\documentclass[printer]{aa}
\usepackage[english]{babel}
\usepackage{graphicx}
\usepackage{newlfont}
\usepackage{amssymb}
\usepackage{subfig}
\usepackage{commath}

\usepackage{amsmath}
\usepackage{latexsym}
\usepackage{amsthm}
\usepackage{natbib}
\usepackage{microtype}
\bibpunct{(}{)}{;}{a}{}{,} 
\usepackage{longtable}
\usepackage{rotating}

\usepackage[usenames,dvipsnames]{color}
\usepackage[bookmarks, colorlinks, breaklinks]{hyperref}  
\hypersetup{linkcolor=blue,citecolor=MidnightBlue,filecolor=black,urlcolor=MidnightBlue} 

\usepackage[varg]{txfonts}

\newcommand{\seciac}[1]{Sect. \ref{#1}}
\newcommand{\figiac}[1]{Fig. \ref{#1}}

\newfont{\gwpfont}{cmssq8 scaled 1000}
\newcommand{\rexcess}{{\gwpfont REXCESS}}

\def\YX {Y_{\textrm X}}
\def\YSZ {Y_{\textrm SZ}}

\def\Mv {M_{500}}
\def\Mv {M_{500}}
\def\Rv {R_{500}}
\def\Rvyx {R_{500}^\mathrm{Y_{X}}}
\def\Mvyx {M_{500}^\mathrm{Y_{X}}}
\def\Mvysz {M_{500}^\mathrm{Y_{\textrm SZ}}}

\def\Mvhe{M_{500}^\mathrm{HE}}

\def\Mvher500{M^\mathrm{HE}}

\def\MY {$M_{500}$--$Y_{\textrm X}$}
\def\MYSZ {$\Mv$--$\YSZ$}

\def\S {$S_{500/2500}$}
\def\csb {$\mathrm{C_{SB}}$}

\def\w{$\langle w \rangle$}

\def\xmm{XMM-{\it Newton}}
\def\planck{{\it Planck}}
\def\chandra{{\it Chandra}}
\usepackage{color}

\def\msol{10^{14}\, {\mathrm M}_{\odot}}

\def\subesz{Low-z PXI}

\def\esz{ESZ}
\def\highz{High-z}
\def\esz{Low-z}
\def\cm{\textit{c}-$M$}

\begin{document}

\title{The Most Massive galaxy Clusters (M2C) across cosmic time: link between radial  total mass  distribution and  dynamical state}

\author{I. Bartalucci\inst{1} \and M. Arnaud\inst{1} \and G. W. Pratt\inst{1} \and J. Démocl\`es\inst{1} \and L. Lovisari\inst{2}}
\institute{AIM, CEA, CNRS, Université Paris-Saclay, Université Paris Diderot, Sorbonne Paris Cité, F-91191 Gif-sur-Yvette, France    \and
Center for Astrophysics | Harvard \& Smithsonian, 60 Garden Street, Cambridge, MA 02138, USA
}

\abstract{We study the dynamical state and the integrated total mass profiles of $75$ massive ($\Mv>5 \times \msol$) Sunyaev-Zeldovich(SZ)-selected clusters at $0.08 < z <1.1$. The sample is built from the \planck\ catalogue, with the addition of four SPT clusters at $z>0.9$.  Using \xmm\ imaging observations, we characterise the dynamical state with the centroid shift $\langle w \rangle$, the concentration \csb, and their combination, $M$, which simultaneously probes the core and the large-scale gas morphology. 
Using spatially resolved spectroscopy and assuming hydrostatic equilibrium, we derive the total integrated mass profiles. The mass profile shape is quantified by the sparsity, that is the ratio of $\Mv$ to $M_{2500}$,  the masses at density contrasts of 500 and 2500, respectively. 
We study the correlations between the various parameters and their dependence on redshift. 
We confirm that  SZ-selected samples, thought to most accurately reflect the underlying cluster population, are dominated by disturbed and non-cool core objects at all redshifts. 
 There is no significant evolution or mass dependence of either the cool core fraction or the centroid shift parameter. The  $M$ parameter evolves slightly with $z$, having a correlation coefficient of   $\rho=-0.2 \pm 0.1$ and a null hypothesis $p$-value of $0.01$.
In the high-mass regime considered here, the sparsity evolves minimally with redshift, increasing by $10\%$ between $z<0.2$ and $z>0.55$, an effect that is significant at less than  $2\sigma$.  In contrast, the dependence of the sparsity on dynamical state  is much stronger, increasing by a factor of  $\sim 60\%$ from the one third most relaxed  to the one third most disturbed objects,  an effect that is significant at more than $3\sigma$.
This is the first observational evidence that the shape of the integrated total mass profile in massive clusters is principally governed by the dynamical state and is only mildly dependent on redshift.  We discuss the consequences for the comparison between observations and theoretical predictions.  }
 \titlerunning{---}
\authorrunning{Bartalucci et al.}

\maketitle

\section{Introduction}\label{sec:introduction}
The shape of the dark matter profile in galaxy clusters  is a sensitive test of the nature of dark matter and of the theoretical scenario of structure formation. 
In the standard  framework, cosmological structures form hierarchically from initial density fluctuations that grow under  the influence of gravity. 
In a $\Lambda$ cold dark matter ($\Lambda$CDM) Universe,  the dark matter (DM) collapse  is scale-free and one expects objects to form with similar internal structure. The DM shape is expected to depend on the halo assembly history, which is a function of the redshift, the total mass, and the underlying cosmology \citep[e.g.][]{dolag2004,kravtsov2012}. A certain scatter of the shape and break of strict self-similarity are expected, reflecting the detailed formation history of each halo.

Clusters of galaxies are ideal targets to test the above scenario: the dark matter is the dominant component by far except in the very centre. Furthermore, complementary techniques can measure the total mass density profile, for example, galaxy velocities; strong gravitational lensing in the centre and weak lensing at large scale; X-ray estimates using the intra-cluster medium (ICM) density and temperature profiles and the hydrostatic equilibrium (HE) equation \citep[see][for a review]{pratt2019}. 

Numerical simulations indeed predict that the cold dark matter density profiles of virialised objects follows a `universal' form. Well-known parametric models for the DM density profiles include those proposed by \citet[herafter, NFW]{nfw1997} and the Einasto profile \citep{ein65}, which is currently considered to be a more accurate description of the profiles in state-of-the-art simulations \citep{nfw04,wu2013}.

A fundamental parameter of these parametric models is their concentration, which in general terms describes the relative distribution in the core as compared to the outer region \citep{klypin2016}. The concentration  is characterised in the NFW and Einasto\footnote{In this model,  the actual shape also depends on a second parameter.} models by the ratio \textit{c}$\equiv r_{-2} / R_{\Delta}$\footnote{$R_{\Delta}$ is defined as the radius enclosing $\Delta$ times the critical matter density at the cluster redshift. $M_{\Delta}$ is the corresponding mass.}, where $r_{-2}$ is the radius at which the logarithmic density slope is equal to $-2$.  The relation between the concentration and the total mass (hereafter \cm) has been very widely used as an indicator of the dark matter profile shape in cosmological simulations \citep[e.g.][and references therein]{die15}.  However, it  has been shown that the \cm\ relation of the most relaxed haloes is different to that derived for the full population \citep[e.g.][]{neto2007,bhattacharya2013}. In fact, the capability of these  parametric models to reproduce the DM profile (i.e. the goodness of the fit) depends on the dynamical state of the halo and on the formation time \citep[][for the NFW case]{jing2000,wu2013}.  Results on the dependence of the DM profile shape on mass and redshift based on \cm\ relations may thus be ambiguous, and depend on the sample selection \citep[][]{klypin2016,balmes2014}. 
For this reason, recent works based on simulations have often focused on relaxed haloes \citep[e.g.][]{dutton2014,ludlow2014,correa2015c}.

However, a rigorous comparison with cluster observations cannot be made, as there is a continuous distribution of dynamical states, and the definition of what is a relaxed object is therefore somewhat arbitrary. Moreover, in the absence of fully realistic simulations including the complex baryonic physics, it is nearly impossible to define  common criteria  that quantify  the dynamical state in a consistent manner,  both in simulations and observations. Ideally we would compare the full cluster population from numerical simulations to observed samples chosen to reflect as closely as possible  the true underlying population.
The advent of Sunyaev-Zeldovich (SZ)-selected cluster catalogues offers a unique opportunity to build such observational samples. Surveys such as those from the Atacama Cosmology Telescope (ACT, \citealt{marriage2011}), the South Pole Telescope (SPT, \citealt{spt1} and \citealt{spt2}), and the \planck\ Surveyor  (\citealt{planckesz}, \citealt{PSZ1}, and \citealt{PSZ2}) have provided SZ-selected cluster samples up to $z\sim 1.5$. The magnitude of the SZ effect being closely linked to the underlying mass with small scatter \citep{das04}, these are thought to be as near as possible to being mass selected, and as such unbiased. 

The observational study of the DM profile shape using such samples requires investigation of the dependence on fundamental cluster quantities such as mass, dynamical state, and redshift. X-ray observations are an excellent tool with which to undertake such studies -- the ICM morphology can be used to infer the dynamical state, while the total mass profile can be derived by applying the hydrostatic equilibrium (HE) equation to spatially resolved density and temperature profiles.
While this method yields the highest statistical precision on individual profiles over a wide radial range and up to high $z$ \citep{amodeo2016,bartalucci18}, it has the drawback of a systematic uncertainty due to any departure of the gas from HE, which must be taken into account. 

Here we apply the sparsity parameter  introduced by \cite{balmes2014}  to quantify the shape of the total mass  profiles derived from the X-ray observations. The sparsity is defined as the ratio of the integrated mass  at two overdensities. This non-parametric quantity is capable of efficiently characterising the profile shape, as long as the two overdensities are separated enough to probe the shape of the mass profile \citep{balmes2014,corasaniti2018}. Formally, the DM profiles can be derived by subtracting the gas and galaxy distribution from the total. 
However in the following we focus on the total distribution, in view of the negligible impact of the baryonic component outside the very central region on the total profile shape for the halo masses and density contrasts under consideration \citep[e.g.][Fig.~2]{velliscig2014}.

In this work, we present the dynamical properties and the individual spatially resolved radial total mass profiles of a sample of 75 SZ-selected massive clusters in the $[0.08$--$1]$ redshift range with $\Mv = [5$--$20] \times 10^{14} M_{\odot}$. 
We discuss the dispersion and evolution of the total mass profile shape, and demonstrate the link between the diversity in profile shape and the underlying dynamical state. In Sect. 2 we present the sample; in Sects. 3 and 4 we describe the methodology used to derive the HE mass profiles and the morphological parameters of each cluster, respectively; in Sect. 5 we discuss the morphological properties of the sample and its evolution; in Sect. 6 we investigate the dependence of the  profile shape on mass, redshift, and dynamical state using the sparsity; and  finally, in Sects. 7 and Sect. 8 we discuss our results and present our conclusions.

We adopt a flat $\Lambda$-cold dark matter cosmology with $\Omega_m = 0.3$, $\Omega_\Lambda = 0.7$, $H_{0} = 70$ km Mpc s$^{-1}$, and $h(z) = (\Omega_m (1+z)^3 + \Omega_\Lambda)^{1/2}$, where $h(z) = H(z)/H_{0}$ throughout. Uncertainties are given at the 68 \% confidence level ($1\sigma$). All fits were performed via $\chi^2$ minimisation.

\section{The sample}\label{sec:data_sample}
\subsection{The high-z SZ-selected sample}

\begin{figure}[]
\begin{center}
\includegraphics[width=1.\columnwidth]{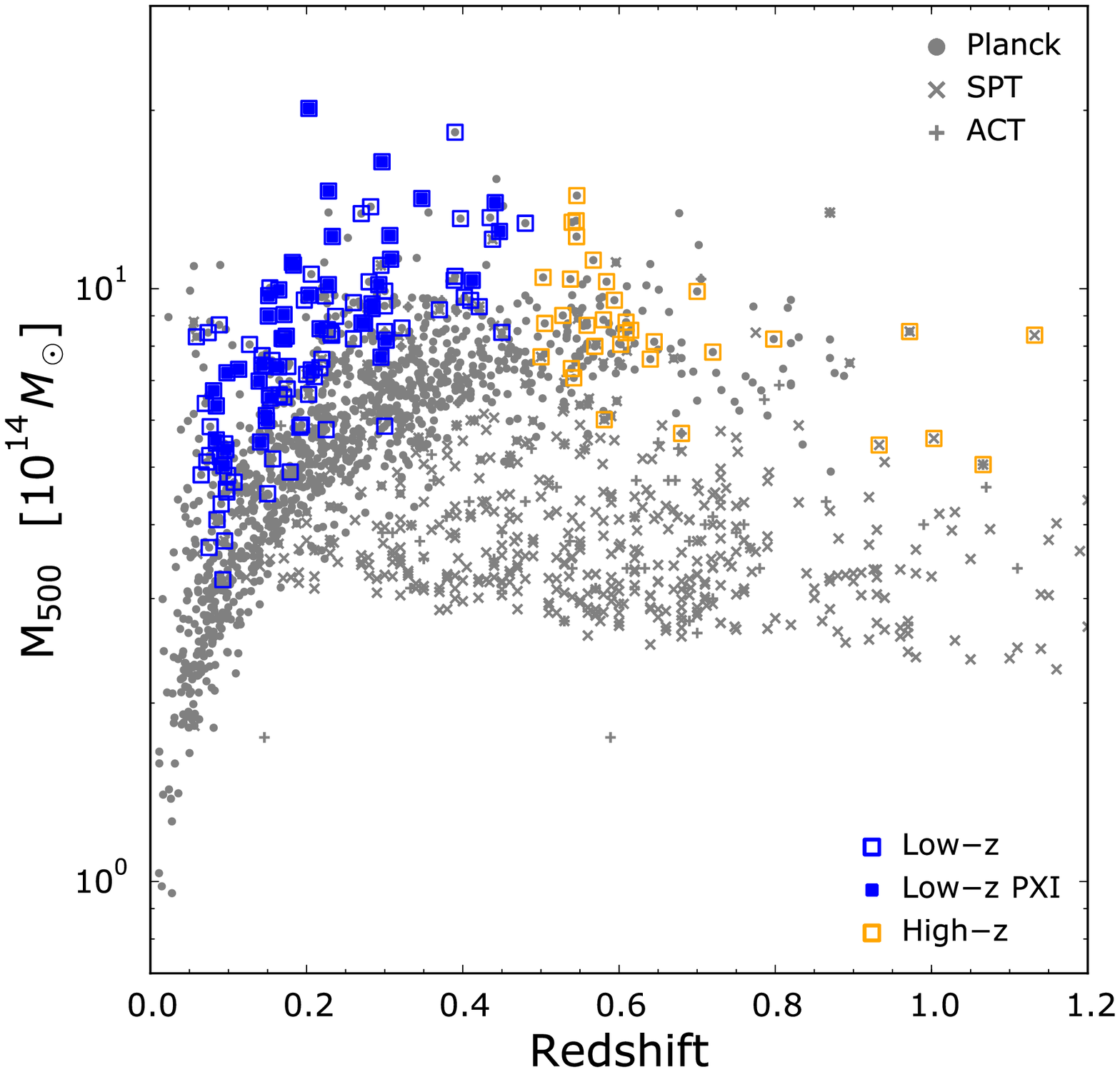} 
\end{center}
\caption{\footnotesize{ Distribution in the mass--redshift plane of all the clusters published in the \planck, SPT, and ACT catalogues. Filled circles: \planck\ clusters with available redshifts (\citealt{planckesz}, \citealt{PSZ1}, \citealt{PSZ2}); crosses: SPT \citep{spt2}; plus symbols: ACT \citep{hasselfield2013}. Masses in the \planck\ catalogue are derived iteratively from the \MYSZ\ relation calibrated using hydrostatic masses from \xmm; they are not corrected for the hydrostatic equilibrium (HE) bias. In the figure the \planck\ masses are multiplied by a factor of 1.2 (i.e. assuming 20\% bias). The blue and orange open squares identify the \esz\ and \highz\ samples considered in this study (see Sect.~\ref{sec:data_sample}). The clusters that are part of the \subesz\ sample are shown as filled blue squares. }}
\label{fig:planck_mz_plane}
\end{figure}
Our  initial sample  is built from the $28$ clusters with spectroscopic $0.5 < z < 0.9$ in the sample of the \xmm\ Large Programme (LP)  ID069366  (with re-observation of flared targets  in ID72378), consisting of clusters detected at high $S/N$ with \planck, and confirmed by autumn 2011 to be at $z > 0.5$.   The exposure times of these observations were optimised  to allow the determination of spatially resolved radial total mass profiles at least up to $\Rv$.
To extend the redshift coverage to $z\sim 1$  we included the five most massive clusters detected by  SPT and \planck\  in the redshift range $0.9<z<1.1$. Using deep \xmm\ and \chandra\ observations, \cite{bartalucci17} and \cite{bartalucci18} examined the  X-ray properties of these objects and determined ICM and HE total mass profiles up to $\Rv$.
The resulting observation properties are detailed in Table 1 of \cite{bartalucci17}. 

Combining the above, we obtain the full `\highz' sample of 33 objects, with $M_{500}^{SZ} > 5 \times \msol$, shown with orange open squares in \figiac{fig:planck_mz_plane}. The observation details of this sample are detailed in Table \ref{tab:500_prop}.

\subsection{The low-z SZ selected sample}\label{sec:local_sample}
Any evolution study requires a local reference sample 
with similar selection and quality criteria 
to act as an `anchor' to compare to high redshifts. The ongoing  (AO17) \xmm\ heritage program `Witnessing the culmination of structure formation in the Universe' (PIs M. Arnaud and S. Ettori), based on the final \planck\ PSZ2 catalogue,  will serve this function in the future (observations are to be completed by 2021). In the interim,  for  the  present  study we use published  \xmm\ follow-up of \planck-selected local clusters taken from the ESZ sample. 

The early SZ  (ESZ, \citealt{planckesz}) catalogue represents the first release derived from the  \planck\  all-sky SZ survey, containing $188$ clusters mostly at $z<0.3$. There are a number of studies in the literature describing the X-ray properties of this sample. In particular, \cite{lov17} characterised the global properties and the morphological state of the ESZ clusters covered by \xmm\ observations. We use their results on the morphological properties of the $118$ ESZ clusters at $0.05<z<0.5$ for which $\Rv$ is within the field of view, and the relative error on the morphological parameters is less than $50\%$. This is about $80\%$ of the corresponding parent ESZ subsample defined with the same $z$ and size criteria, so we do not expect any major bias due to the incomplete \xmm\ coverage. These objects cover one decade in mass, $M_{500} \sim [3-20] \times \msol$, and are shown with blue open squares in \figiac{fig:planck_mz_plane}. Henceforth  we refer to this sample as the `\esz' sample. 

The spatially resolved thermodynamic properties of an ESZ subsample were analysed by the \planck\ Collaboration, who combined X-ray and SZ data to calibrate the local scaling relations \citep{pipxi} and measure the pressure profiles \citep{planck_pressure}. We use the published thermodynamic profiles of $42$ clusters for which $\Rv$ is within the field of view (the subsample "A" defined in Sect.~3.1 of \citealt{pipxi}) to derive the total mass profiles. These clusters are shown as filled blue squares in \figiac{fig:planck_mz_plane}. We henceforth refer to this sample as the "\subesz" sample. Its representativeness with respect to the full \esz\ sample is excellent, as discussed in Appendix \ref{sec:esz_repr}. 

\subsection{Data preparation}
The observations used in this work were taken using the European Photon Imaging Camera (EPIC, \citealt{turner2001} and \citealt{struder2001}) instrument on board the \xmm\ satellite. This instrument is composed of three CCD arrays, namely MOS1, MOS2, and PN, which simultaneously observe the target. Datasets were reprocessed using the Science Analysis System\footnote{cosmos.esa.int/web/xmm-newton} (SAS) pipeline version $15.0$ and calibration files as available in December $2016$. Event files with this calibration applied were produced using the \verb?emchain? and \verb?epchain? tools. 

The reduced datasets were filtered in the standard fashion. 
Events for which the keyword PATTERN is $<4$ and $<13$ for MOS$1,2$, and PN cameras, respectively, were filtered out from the analysis. Flares were removed by extracting a light curve, and removing from the analysis the time intervals where the count rate exceeded $3\sigma$ times the mean value. We created the exposure map for each camera using the SAS tool \verb?eexpmap?.
We merged multiple observation datasets of the same object, if available. We report in Table \ref{tab:500_prop} the effective exposure times after all these procedures. We corrected vignetting following the weighting scheme detailed in \cite{arnaud2001}. The weight for each event was computed by running the SAS \verb?evigwieght? tool on the filtered observation datasets.

Point sources were identified using the Multi-resolution wavelet software \citep{sta98} on the exposure-corrected $[0.3-2]$ keV and $[2-5]$ keV images. We inspected each resulting list by eye to check for false detections and missed sources. We defined a circular region around each detected point source 
and excised these from the subsequent analysis. 

We also defined regions encircling obvious sub-structures. Identified by eye, these regions were considered in the morphological analysis because the parameters we used to characterise the morphological state of a cluster (\seciac{sec:morpho_analysis}) are sensitive to the presence of any such sub-structure. However, these regions were excluded in the radial profile analysis 
detailed in Sect.~\ref{sec:radial_profiles}.

X-ray observations are affected by instrumental and sky backgrounds. The former is due to the interaction of the instrument with energetic particles, while the latter is caused by Galactic thermal emission and the superimposed emission of all the unresolved point sources, namely the cosmic X-ray background  \citep{lumb2002, kuntzsnowden2000}. These components were estimated differently for 
the radial profile 1D analysis and for the morphological 2D analysis, as described in Sects.~\ref{sec:radial_profiles} and \ref{sec:morpho_analysis}, respectively.
\section{Radial profile analysis}\label{sec:radial_profiles}
\subsection{Instrumental background estimation}

We evaluated the instrumental background for the radial profile analysis following the procedures described in \cite{pratt2010}. Briefly, observations taken with the filter wheel in CLOSED position  
were renormalised to the source observation count rate in the $[10-12]$ and $[12-14]$ keV bands for the EMOS and PN cameras, respectively. We then projected these event lists in sky coordinates to match our observations. We applied the same point source masking and vignetting correction to the CLOSED event lists as for the source data. We also produced event lists to estimate the out-of-time (OOT) events using the SAS-\verb?epchain? tool. 

\subsection{Density and 3D temperature profiles}

To determine the radial profiles of density and temperature of the ICM we followed the same procedures and settings detailed in Sect. 3.2 and Sect. 3.3 of \cite{bartalucci17}. 
Briefly, we firstly determined the X-ray peak by identifying the peak of the emission measured in count-rate images in the [0.3-2.5] keV band smoothed using a Gaussian kernel with a width of between 3 and 5 pixels. We extracted the vignetted-corrected and background-subtracted surface brightness profiles, $S_X$, from concentric annuli of width $2\arcsec$, centred on the X-ray peak, from both source and background event lists. These profiles were used to derive the radial density profiles, $n_e (r)$, employing the deprojection and PSF correction with regularisation technique described in \cite{croston2006}.

We obtained the deprojected temperature profiles by performing the spectral analysis described in detail in \citet{pratt2010} and Sect. $3.4$ of  \cite{bartalucci17}. The background-subtracted spectrum of a region free of cluster emission was  fitted with two unabsorbed {\sc MeKaL} thermal models plus an absorbed power law with fixed slope of $\Gamma=1.4$. The resulting best-fitting model, renormalised by the ratio of the extraction areas, was then added as an extra component in each annular fit. The cluster 
emission was modelled by an absorbed {\sc MeKaL} model with $N_{\mathrm H}$ given in Table~\ref{tab:500_prop} using the absorption cross sections from  \citet{mmc83}.
Spectral fitting was performed using XSPEC\footnote{\url{https://heasarc.gsfc.nasa.gov/xanadu/xspec/}} version 12.8.2.  The deprojected 3D temperature profile, $T_\mathrm{3D}$, was derived from the projected profile using the "Non parametric-like" technique described in Sect.~2.3.2 of \cite{bartalucci18}.

For the \subesz\ sample, we used the temperature and density profiles published in \cite{planck_pressure} and \citet{pipxi},  which were derived employing identical methods to those used in this work. 

\subsection{Global properties}
We determined the mass at density contrast $\Delta=500$, $\Mvyx$, and corresponding  $\Rvyx$ radius,  from the mass proxy  $\YX$. This was computed iteratively from the \MY\ relation, calibrated by \cite{arnaud2010} using HE mass estimates of local relaxed clusters. We assumed that  the \MY\ relation obeys self-similar evolution. The starting mass value was obtained from the $M-T$ relation of \cite{arnaud2005}, and the computation converges typically within 5-10 iterations. The quantity $\YX$ is defined as the product of the temperature measured in the $[0.15-0.75]\Rv$ region and the gas mass within $\Rv$ \citep{krav2006}, the gas mass profiles being computed from the density profiles. The mass for each cluster and associated errors are reported in Table \ref{tab:500_prop}.

We used the  $\Mvyx$ published  by \cite{lov17} for the \esz\ sample, 
and the $\Mvyx$ computed by us for the \subesz\ and \highz\ samples. We investigated the coherence between these measurements  by comparing the masses for the clusters in common between the Low-z plus High-z samples and the \subesz\ sample. The excellent agreement between the two is discussed in Appendix \ref{sec:ll_vs_jd}, and the comparison is shown the right panel of Fig.~\ref{fig:jd_vs_ll}.

\subsection{Derivation of the total mass profiles}

Under the assumption that the ICM is in hydrostatic equilibrium in  the gravitational potential well the relation between the total halo mass within radius $R$  and the ICM thermodynamic properties is:
\begin{equation}\label{eq:hydro}
M(\leq R) =    -\frac{kT(R)\,R}{G\mu m_\mathrm{H}}\left[\frac{\dif\,\ln{n_\mathrm{e}\,(R)}}{\dif\,\ln{R}}+ \frac{\dif\, \ln{T(R)}}{\dif\, \ln{R}}\right],
\end{equation}
where G is the gravitational constant, $m_\mathrm{H}$ is the hydrogen atom mass, and $\mu=0.6$ is the mean molecular weight in atomic mass unit.  We used the deprojected density and temperature profiles and the relation in Eq. \ref{eq:hydro} to derive the mass profiles,  applying the `forward non-parametric-like technique' detailed in Sect.~2.4.1 of \cite{bartalucci18}.  The `forward' in the technique name is due to the fact that we started our analysis from the surface brightness and projected temperature observables to derive the mass profiles at the end. The `non-parametric-like' refers to the fact that we used a deprojection technique to derive the density and temperature profiles, that is, without using parametric models. 

\begin{figure}[!t]
\begin{center}
\includegraphics[width=1.\columnwidth]{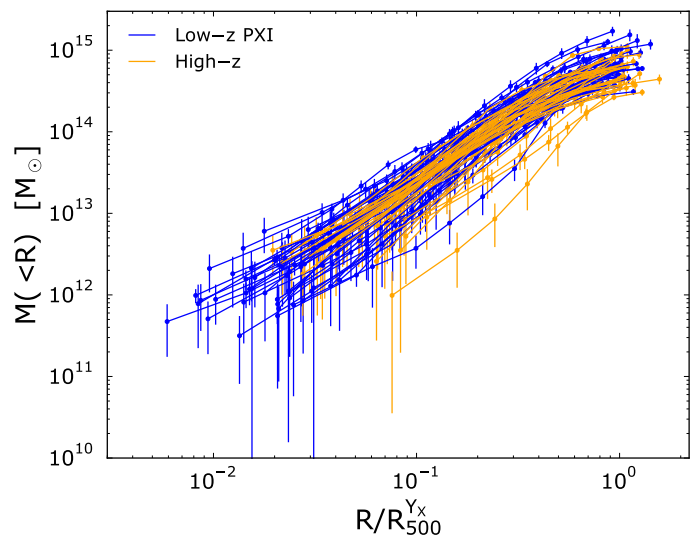} 
\end{center}
\caption{\footnotesize{ Integrated mass profiles as a function of scaled radius, estimated from  the hydrostatic equilibrium equation, for all clusters  considered in this work.  The  \subesz\ and \highz\ samples are plotted in blue and red, respectively. }}
\label{fig:Mprofraw}
\end{figure}

The mass profiles of the \subesz\ and \highz\ samples and their associated uncertainties are shown in Fig. \ref{fig:Mprofraw} (those of the five  highest-redshift clusters are reproduced as published in \citealt{bartalucci18}). The profiles of the \highz\ clusters are mapped at least up to $0.8\,\Rv$, with $20$ out of $33$  objects measured up to $\Rv$, thus not requiring extrapolation to compute the HE mass at a density contrast of $\Delta=500$, $\Mvhe$. The median statistical error is about $20\%$. The quality of the \subesz\ mass profiles, based on archival data, is lower and less homogeneous.  Only $18$ out of $42$ clusters have temperature profiles extending to $\Rv$, with a maximum radius  between $0.6\,\Rv$ and $1.4\,\Rv$.  \cite{bartalucci18} showed that while the HE mass is very robust when the temperature profiles extend up to $\Rv$,  the  $\Mvhe$ mass is very sensitive to the mass estimation method when extrapolation is required. This is  particularly the case  for irregular clusters.  To minimise systematic errors,  we used $\Mvyx$ rather than  $\Mvhe$  in the following for all scaling with mass,  and for the computation of the sparsity (see below). The possible impact of this choice on our results is discussed in Sect.~\ref{sec:discussion}.

\subsection{Sparsity}
\label{sec:spardef}
The sparsity, $S$,  was introduced by \cite{balmes2014}  to quantify  the shape of the dark matter profile. It is defined as the ratio of the integrated mass at two over-densities, and has the advantage of being  non-parametric, as there is no \textit{a priori} assumption on the form of the profile. The sparsity therefore represents a useful measure when dealing with a population of objects with a wide variety of dynamical states. Another non-parametric approach, advocated by \cite{klypin2016}, considers the maximum circular velocity, which is linked to the traditional NFW or Einasto concentration.  As the velocity is basically the square root of $M(<R)/R$,  it can also be derived from observations. In practice however, measuring such a maximum is much more difficult than measuring a ratio of integrated masses.  

In the following, we concentrate on the sparsity to investigate the shape of the mass profiles, which is defined as:
\begin{equation}
S_{\Delta_1,\Delta_2} \equiv  \frac{M_{\Delta_1}}{M_{\Delta_2}},
\end{equation}
where $M_{\Delta}$  is the mass corresponding  to the density contrast $\Delta$ and with $\Delta_1 < \Delta_2$. 
We recall that $M_{\Delta}= M(<R_\Delta)$, which is the mass enclosed within $R_\Delta$,  such that
\begin{equation}
\frac{M(<R_\Delta)}{(4\pi/3)\,R_{\Delta}^3} = \Delta\, \rho_{crit}
\label{eq:mdelta}
.\end{equation}
\cite{balmes2014} argue that the general properties of the sparsity do not depend on the choice of $\Delta_{1,2}$ as long as the halo  is well defined (i.e. $\Delta_1$ is not too small), and that the  interaction between dark and baryonic matter in the central region can be neglected  (i.e. $\Delta_2$ is not too large). We use  $\Delta_1=500$  and 
$\Delta_2=2500$;  the choice of the latter is further discussed in Sect.~\ref{sec:sparobs}.  

%

\section{Morphological analysis}\label{sec:morpho_analysis}

\subsection{Centroid shift}
\label{sec:morphoanal}

We produced count images for each camera in the soft band, $[0.3$--$2]$ keV, binned using $2"$ pixel size, on which we excised and refilled the masked regions where point sources were detected using the \textit{Chandra} interactive analysis of observation (CIAO, \citealt{fruscione2006})
 \verb?dmfilth?  tool. 
Sub-structures were not masked for this analysis. We estimated the background following a similar approach to that of \citet{bohringer2010}. We computed the background map for each camera by fitting the refilled count images using a linear combination of the vignetted and unvignetted exposure maps to account for the instrumental and sky background, respectively. We removed the cluster emission by masking a circular region within $\Rvyx$ and centred on the X-ray peak. Exposure maps and background and count images of MOS1, MOS2, and PN were combined, weighting by the ratio of integrated surface brightness profile of each individual camera to that of the combined profile. The combined count images were then background subtracted and exposure corrected.  We produced 100 realisations of the count-rate maps by applying the same procedure to $100$ Poisson realisations of the count maps. 

The centroid shift parameter, \w, was introduced by \cite{mohr1993} as a proxy to characterise the dynamical state of a cluster. The centroid $(x_c , y_c)$ within an aperture is defined as
\begin{equation}
(x_c , y_c) \equiv \frac{1}{N_i} \sum_k n_k (x_k ; y_k),
\end{equation}
where $N_i$ is the total number of counts per second within the \textit{i}th aperture and $n_k$ is the count rate in the pixel $k$ of coordinates $(x_k ; y_k)$. We computed the mean deviation of the centroid from the X-ray peak by measuring the displacement within $N=10$ apertures using the definition of \cite{bohringer2010} :
                \begin{equation}\label{eq:centroid}
                \langle w \rangle =  \left(  \frac{1}{N-1} \sum_{i=0}^{10}  (\Delta_i - \langle \Delta \rangle )^2 \right)^{1/2}  \frac{1}{\Rvyx},
                \end{equation}
where $\Delta_i$ is the projected distance between the X-ray peak and the centroid computed within the \textit{i}\,th concentric annulus, each one being $i \times 0.1\Rvyx$ in width. 
Uncertainties on the centroid shift were estimated by measuring \w\ on the 100 Poisson count map realisations, and taking the values within $68\%$ of the median. 

\cite{maughan2008} measured \w\  on a sample of clusters observed by \chandra, excluding the inner $30$~kpc to make the parameter less sensitive to very bright cores. The PSF of \xmm\ is larger for all the clusters considered in this work and, for this reason, we did not excise the core from the analysis. The good agreement  between the \w\ values derived at $z>0.9$ by \chandra\  and \xmm\, shown by \citet{bartalucci17} indicates that the \xmm\ PSF is not an issue.
 \cite{nurgaliev2013} showed that the centroid shift can be biased high in the case of observations with a low number  ($<2000$) of counts. In our sample the minimum number of counts in the $[0.3$--$2]$~keV band we used to measure \w\ is $3000$. Furthermore, all clusters are in the high-SN regime, the lowest SN in our sample being $~40$.

\begin{figure*}[!ht]
\begin{center}
\includegraphics[width=1\textwidth]{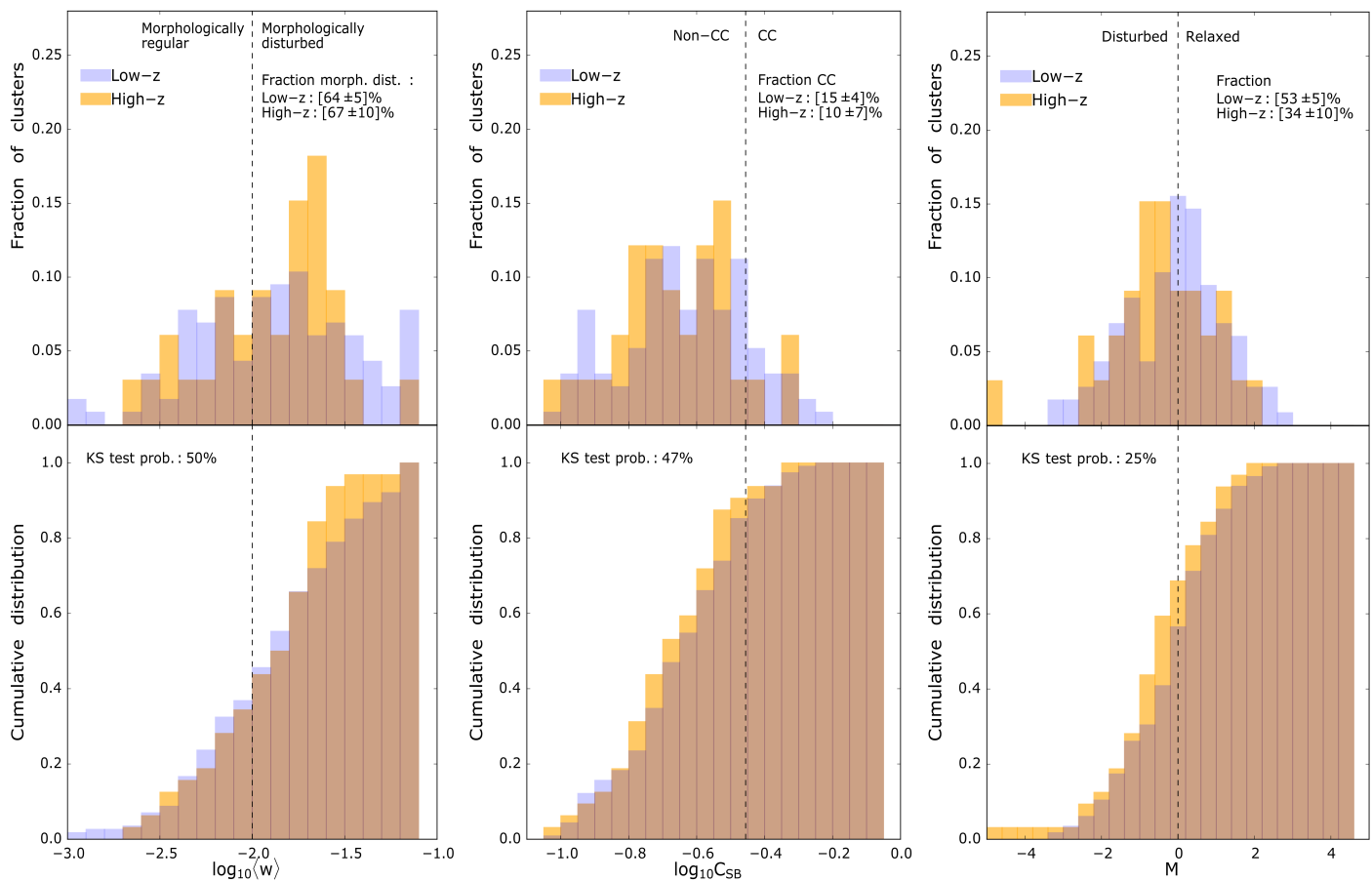} 
\end{center}
\caption{\footnotesize{From left to right, normalised histogram (top panels) and cumulative distributions (bottom panels)  of the  centroid shift, $\langle w\rangle$ ,  the concentration \csb\,, and the $M$ parameter (Eq. \ref{eq:m_ll}). The \csb\ of the \highz\ sample has been $k$-corrected as described in Sect. \ref{sec:csb}.
 The \esz\ and \highz\ distributions are shown in blue and orange, respectively.  The vertical dotted line represents the threshold value  for each parameter. {\it Left panel:} $\langle w \rangle = 0.01$ threshold, separating  morphologically  regular and   disturbed clusters. \textit{Central panels}: \csb = 0.35 threshold, between cool core (CC) and non-CC objects. \textit{Right panels}:  $M=0$ threshold, between disturbed and relaxed objects. The corresponding fraction above each threshold is given in the top right of each  figure. In the bottom  panels, we report  the p-value of  the null hypothesis (i.e. that the two distributions are drawn from the same distribution) from the Kolmogorov-Smirnov  (KS) test.
The distribution of morphologically disturbed and CC objects, based on the parameter \w\ and \csb,  are not statistically different in the \highz\ and \esz\ samples. However, the combination of the two parameters, $M$, indicates that the fraction of disturbed objects is significantly higher in the \highz\ sample. 
}}
\label{fig:morpho_histos}
\end{figure*}

Following \cite{pratt2009}, we initially classify an object as `morphologically disturbed'  if $\langle w \rangle > 0.01$ and `morphologically regular'  if $\langle w \rangle < 0.01$ .  
The results of the centroid shift characterisation for the \esz\ and \highz\ samples is shown in the top- and bottom-left panels of \figiac{fig:morpho_histos}, and the corresponding values are given in Table~\ref{tab:500_prop}.

\subsection{Surface brightness concentration \csb}
\label{sec:csb}
The ratio of the surface brightness profile within two concentric apertures, hereafter  the  \csb, was introduced by \cite{santos2008} to quantify the concentration of cluster X-ray emission.  We computed the \csb\ using the following definition: 
\begin{equation}\label{eq:csb}
\mathrm{C_{SB}} = \frac{\int^{0.1 \times \Rvyx}_0 \mathrm{S_X (r)}\, \mathrm{dr}}{\int^{0.5 \times \Rvyx}_0 \mathrm{S_X (r)}\, \mathrm{dr} },
\end{equation}
where the error was computed using a Monte-Carlo procedure on 100 Gaussian realisations of the surface brightness profiles and taking the $68\%$ value around the median.  The \csb\ parameter is a robust X-ray measurement, as it relies on the extraction of surface brightness profiles only and is not model-dependent. We nonetheless corrected for the \xmm\ PSF in view of the high $z$ and small angular size of the high $z$ sample. 

\cite{santos2010} demonstrated that for objects at high redshift the emission within two apertures requires a different $k$-correction due to the presence of a cool core, that is cool cores will have typically a softer spectrum than the surrounding regions. This correction is potentially important for this study, as we are comparing \csb\ in a wide redshift range. \citet{santos2010} proposed a correction for this effect which requires spatially resolved temperature profiles. At the median redshift of the \esz\ sample, the $k$ correction is negligible ($< 1\%$), but it can be up to $\sim 5\%$ at the highest redshifts. We therefore did not make this correction for the \esz\ sample. For the \highz\ sample, 
  we applied the $k$-correction to the \csb\ of all the clusters as if they were observed at the median redshift of the \esz\ sample, z=0.19. Henceforth, all the \csb\ values shown and used in this work are $k$-corrected in this manner. The results of the \csb\ analysis are reported in the top and bottom central panels of \figiac{fig:morpho_histos} and in  Table \ref{tab:500_prop}.
 
The \csb\ allows the identification of cool-core (hereafter CC) clusters, the parameter being tightly correlated with the cooling time  \citep[e.g.][]{croston2008, santos2010,pascut2015}. 
From the correlation between central density and cooling time in the \rexcess\ sample,  \cite{pratt2009} defined a central density of $n_\mathrm{e,0}=0.04\,{\mathrm cm}^{-3} h(z)$  as a threshold which segregates CC and non-CC clusters (their Fig.~2). We computed the central density for the \subesz\ and \highz\ objects and used the correlation  between this quantity and the \csb\ to translate this density threshold in terms of the \csb , finding  that CC clusters have \csb$>0.35$ for this classification scheme. 

The \csb\ value can also be used as an indicator of the relaxation state of the cluster. A high concentration is an indication  that the core has not been disturbed by recent  merger events. The corresponding threshold defined to distinguish relaxed clusters, for example from the anti-correlation  observed between \w\ and \csb\  and/or visual inspection \citep{cassano2010,lov17}   differs from that used to define the CC/NCC segregation.

 \subsection{Combined dynamical indicator, $M$}
The combination of certain morphological parameters has been shown to identify the most disturbed and relaxed clusters \citep{cassano2010,rasia2013,lov17,cialone2018}.  We take advantage  of the observed anti-correlation between the centroid shift and the \csb\  to compute the  $M$ parameter introduced by \cite{rasia2013} and use it as an additional dynamical indicator. M is  defined as follows.
\begin{equation}\label{eq:m_ll}
M \equiv \frac{1}{2} \left(    \frac{C_\mathrm{SB} - C_\mathrm{SB, med}}{| C_\mathrm{SB, quar} - C_\mathrm{SB, med}|} - \frac{ \langle w \rangle - \langle w \rangle_\mathrm{med}   }{| \langle w \rangle_\mathrm{quar}  - \langle w \rangle_\mathrm{med} |}     \right),
\end{equation}
where $C_\mathrm {SB, med}$ and $\langle w \rangle_\mathrm{med}$ are the median values of the \csb\ and centroid shift, respectively, and $C_\mathrm{SB, quar}$ and $ \langle w_\mathrm{quar} \rangle$ are the first or third quartile depending on whether the  parameter value  is larger or smaller than the median, respectively. The  $M$ parameter is therefore an indicator that combines the large-scale (i.e. centroid shift) and the core (i.e. concentration) properties. 
It is interesting to note that the two morphological parameters appear in Eq. 7 with the same weight to distinguish relaxed and disturbed objects.
This is consistent with what has been derived by \citealt{cialone2018}.
According to this definition, clusters which are characterised by the presence of a cool core and are morphologically regular will have $M>0$ and very disturbed objects with a very diffuse core will have $M<0$. 
\begin{figure*}[!t]
\begin{center}
\resizebox{\textwidth}{!} {
\includegraphics[]{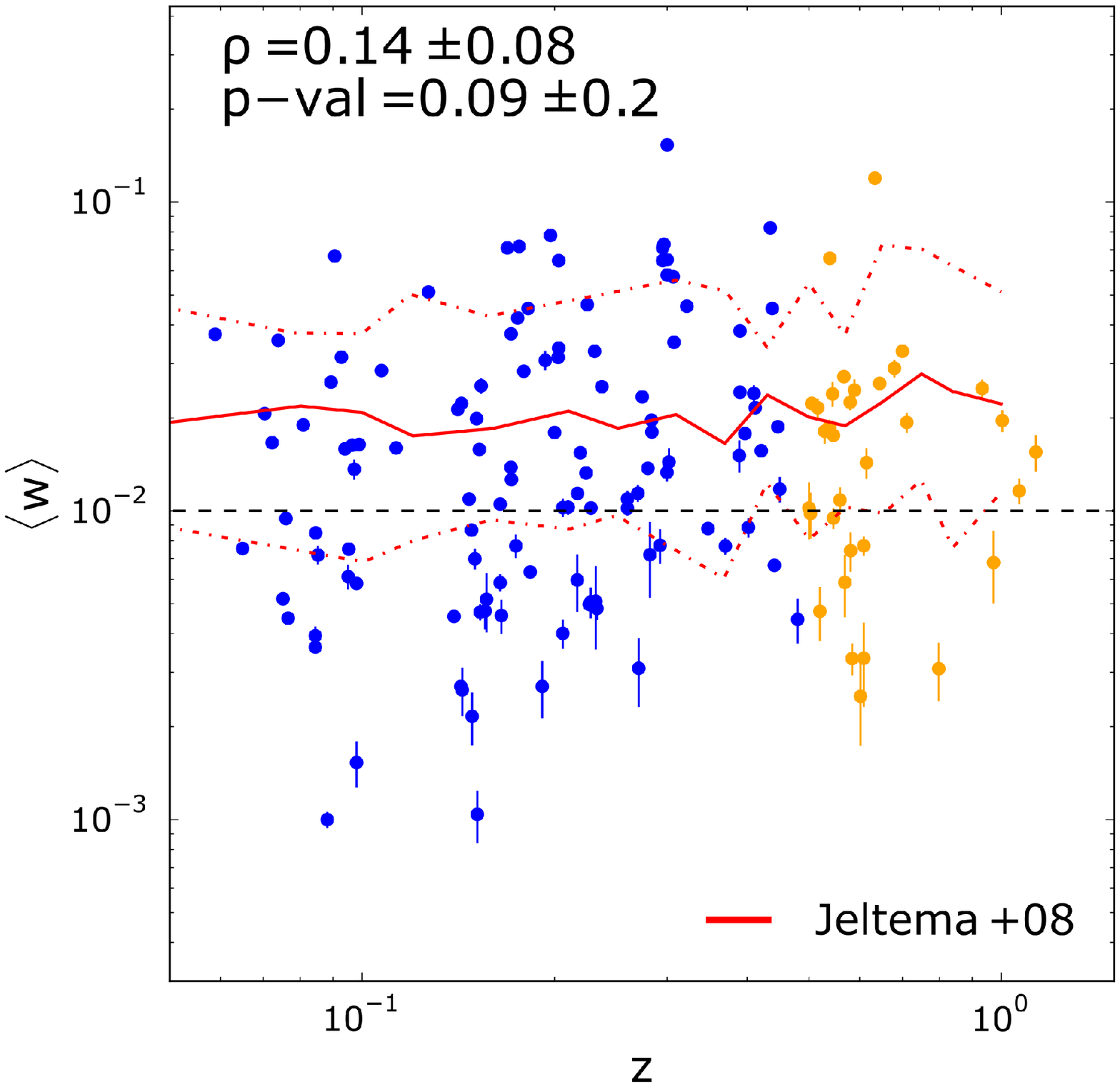}
\includegraphics[]{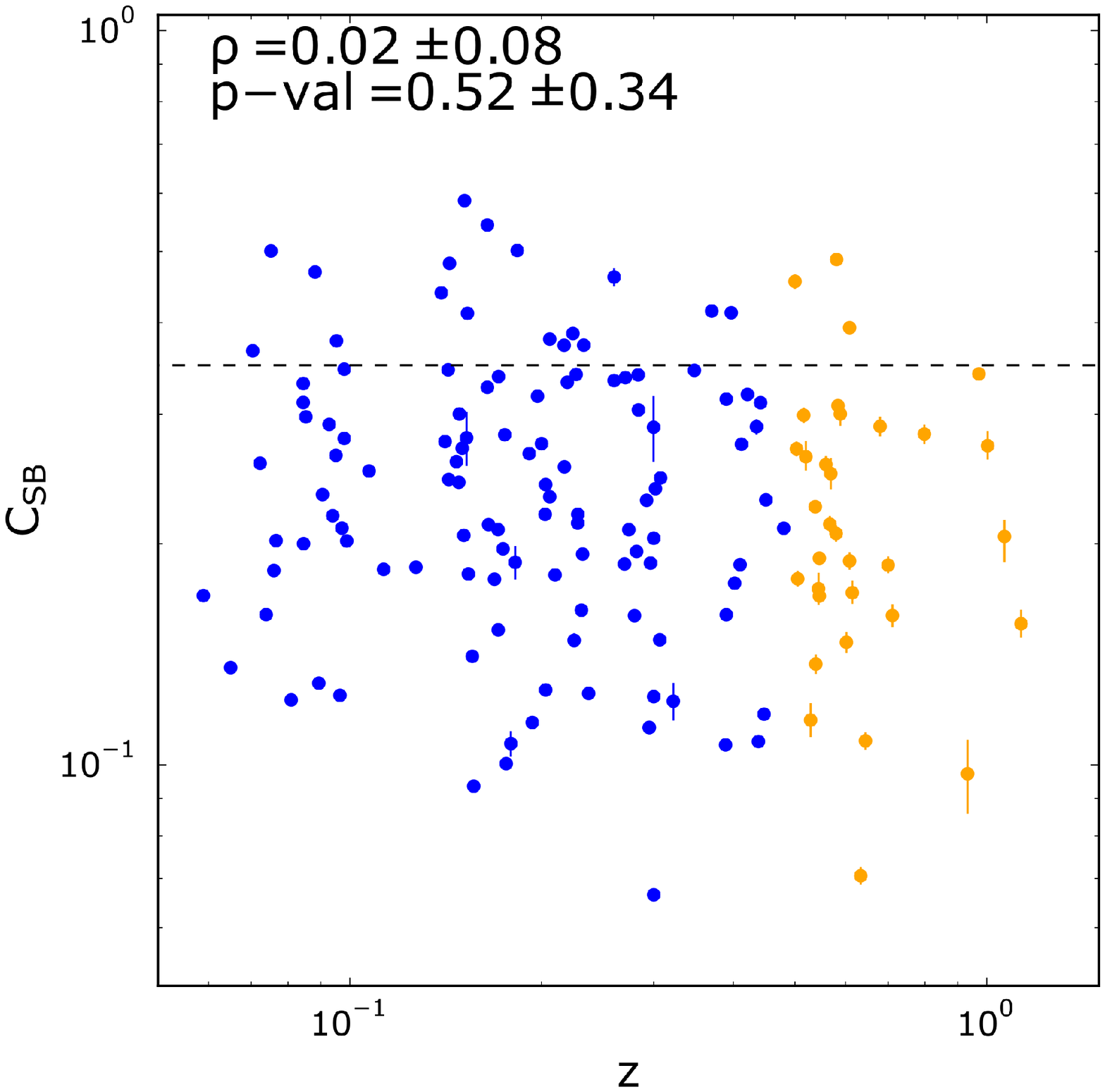}
\includegraphics[]{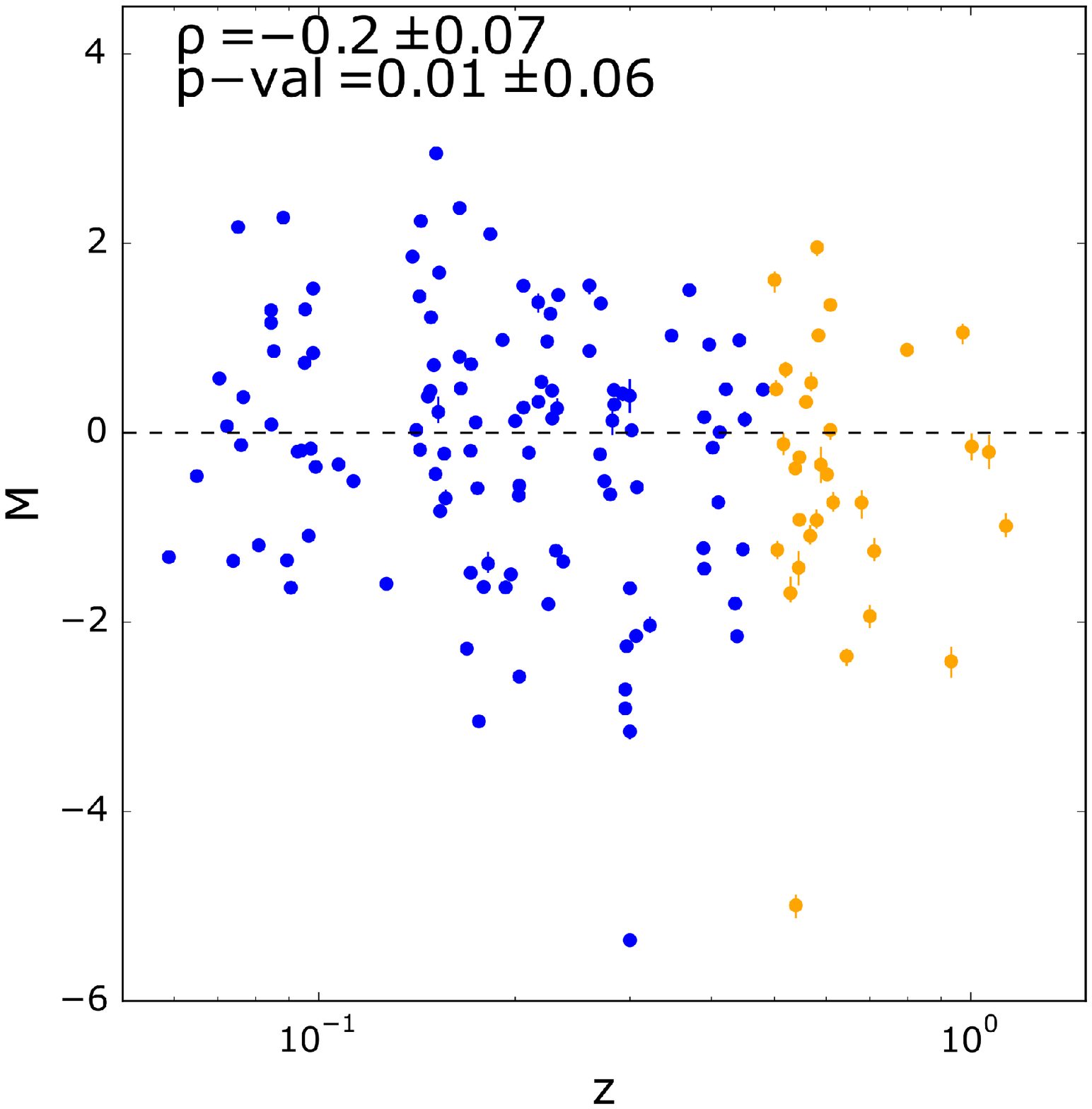}
}
\resizebox{\textwidth}{!} {
\includegraphics[]{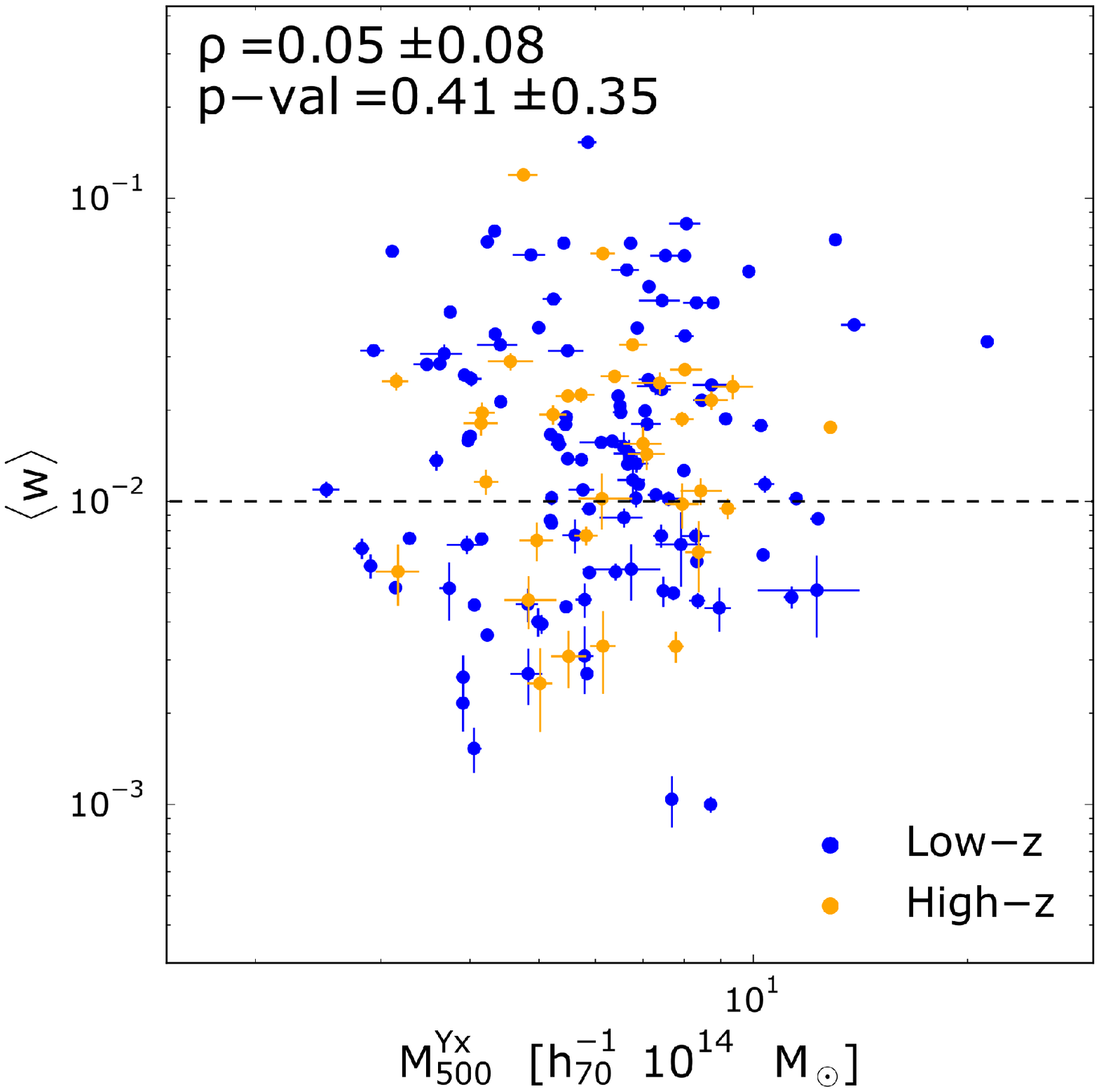}
\includegraphics[]{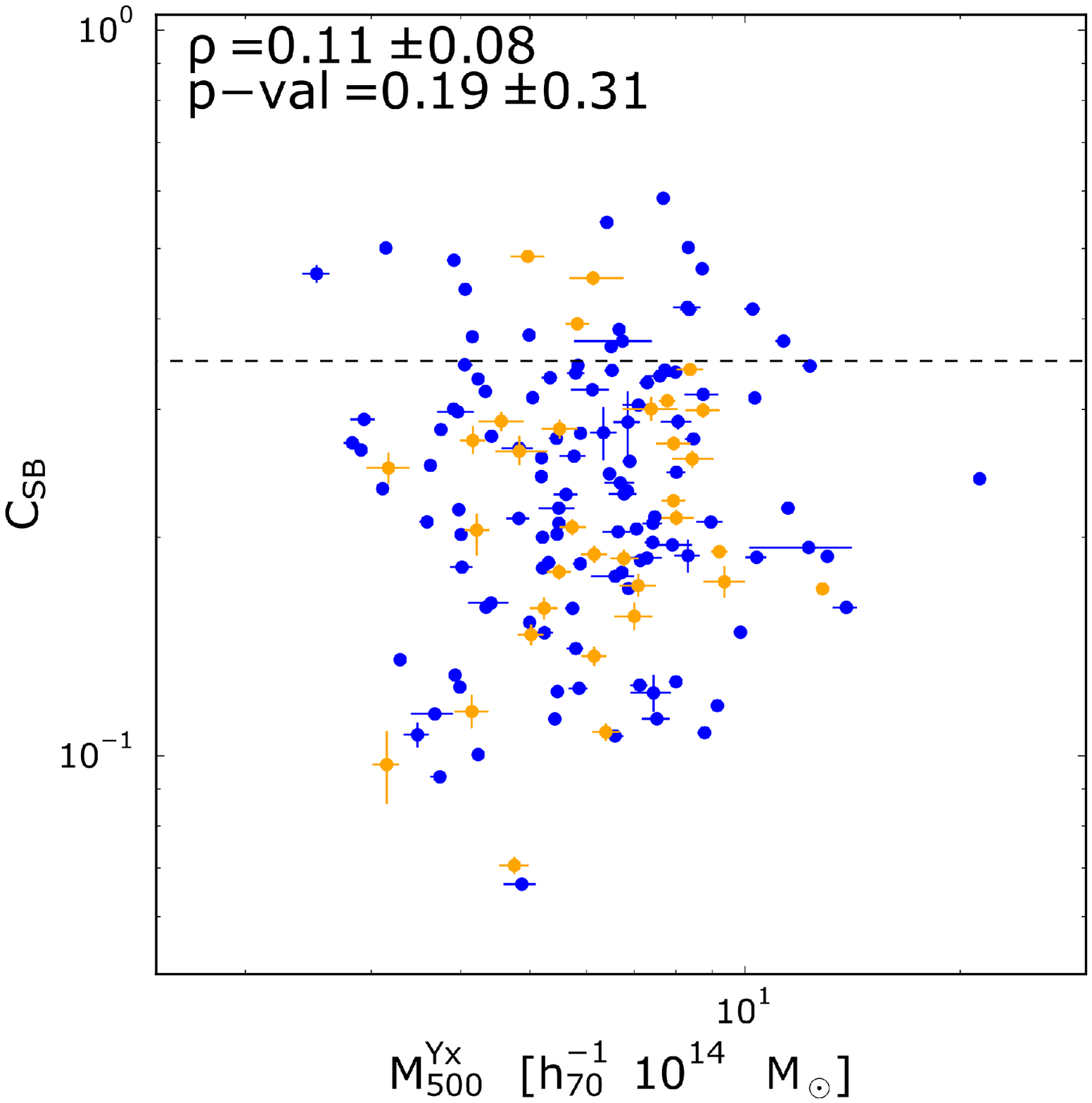}
\includegraphics[]{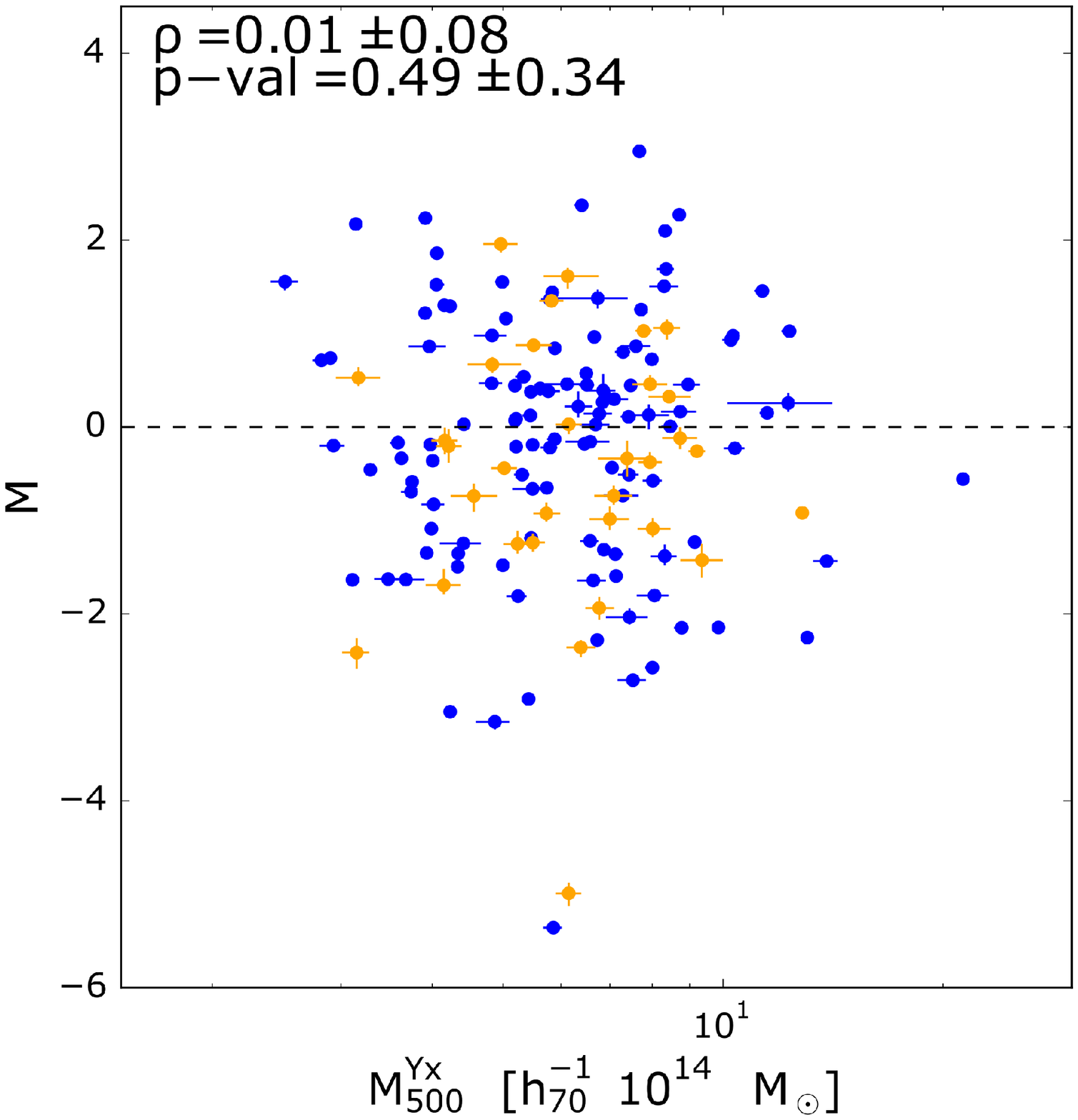}
}
\end{center}
\caption{\footnotesize{Morphological parameters  vs.  redshift  (top panel) and mass (bottom panel) of all the clusters used in this work. \highz\ and \esz\ sample clusters are colour-coded in blue and orange, respectively, following the sample colour code  of  \figiac{fig:morpho_histos}.  \textit{Left panels}:  Centroid shift, \w.  The solid red line is  the mean \w\  derived by \cite{jeltema2008} from numerical simulations. The dotted lines correspond  to  the $\pm68\%$ dispersion,   \textit{Middle panel:} Concentration, \csb. \textit{Right panel: } $M$ parameter. In each panel, the horizontal dotted line  identifies the corresponding threshold,  as defined in  \figiac{fig:morpho_histos}. The symbols   $\rho$ and p-val correspond to the Spearman's rank correlation factor and the corresponding null hypothesis $p$-value, respectively. Errors on these quantities are computed through 1000 bootstrap resampling. 
The figure corroborates the lack of evolution with redshift of \w\ and \csb\ shown in \figiac{fig:morpho_histos}.  However, there is a  mild but significant evolution of the combined parameter $M$. There is no dependence on mass of the morphological parameters.
}}
\label{fig:morpho_evo}
\end{figure*}
Henceforth, we  refer to the former and latter objects as `relaxed' and `disturbed', respectively. 
The choice of this dual classification is arbitrary and does not correspond to a strict segregation between two types of objects. The distribution of $M$ both for \esz\ and \highz\ samples is continuous, as shown in the top right panel of \figiac{fig:morpho_histos}. The results of the $M$ characterisation are reported in the top- and bottom-right panels of \figiac{fig:morpho_histos} and in Table~\ref{tab:500_prop}.
We note that the numerical value of $C_\mathrm{SB, med}$ that we use is smaller than the threshold used in Sect.~\ref{sec:csb} to define CC clusters, and is closer to the value chosen  by \citet{lov17} after visual classification  of clusters  when defining a similar M parameter \footnote{ $C_\mathrm{SB, med}=0.23$ which is close  to $C_\mathrm{SB}=0.15$, used by \citet{lov17}, after correction for  the different aperture definition (their Fig.~C1) }.  

\subsection{Consistency of the  morphological characterisation}
We used the results of the morphological analysis published in \cite{lov17} to characterise the \esz\ cluster morphological properties, using their \w\ and \csb\ values to derive the $M$ parameter. We obtained the  morphological parameters of the \highz\ sample using a different pipeline and different analysis settings. To avoid potentially biased conclusions on morphological evolution or  the dependence of the  mass profiles on morphology for the full  sample, it is necessary to check the consistency between our morphological analysis and that of \cite{lov17}.  We thus compared the morphological parameters derived from each pipeline independently for the common clusters of the \subesz\ sample. The agreement is  excellent, as shown in the left and central panel of Fig.~\ref{fig:jd_vs_ll}. Full details of the comparison are discussed in Appendix~\ref{sec:ll_vs_jd}.

\begin{figure*}[!t]
\begin{center}
\includegraphics[width=\textwidth]{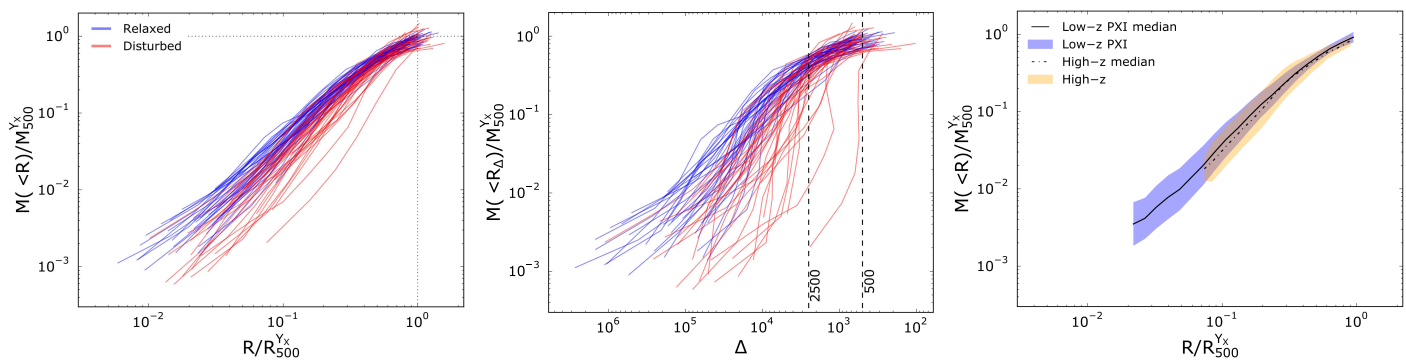} 
\end{center}
\caption{\footnotesize{ Scaled radial mass profiles extracted assuming hydrostatic equilibrium for the samples considered in this work. \textit{Left panel:} Radius and mass are scaled by $\Rvyx$ and $\Mvyx$, respectively. 
 Blue and red profiles represent relaxed  and disturbed  objects, respectively, according to the $M$ parameter. 
 \textit{Central panel:} Same as the left panel, except for the fact that we show the mass profile as a function of the density contrast, $\Delta$. The vertical line $\Delta = 2500$ is the overdensity used in this work to compute the sparsity described in Sect. \ref{sec:sparsity}. \textit{Right panel:} Comparison of the scaled mass profiles for the \subesz\ and \highz\ samples.
 The solid line and dotted lines represent the median computed for the \subesz\ and \highz\ samples, respectively. The gold and green shaded regions represent the $1\sigma$ dispersion of the \subesz\ and \highz\ samples, respectively. 
 Disturbed clusters have a shallower mass distribution, and present a larger dispersion than that of the most relaxed objects. On the other hand, the median HE mass profile depends mildly on  redshift.
  }}
\label{fig:he_profiles}
\end{figure*}
\section{Morphology and dynamical state}
\subsection{Sample characterisation and  comparison }\label{sec:chara_morpho}

The results of the \esz\ and \highz\ morphological characterisation are shown in \figiac{fig:morpho_histos}, where the top and bottom of each panel show the normalised and cumulative distribution for each parameter, respectively.  We also show the fraction of objects above the fiducial thresholds discussed in Sect.~\ref{sec:morphoanal} in each panel. Errors are computed from 1000 Monte-Carlo bootstrap resamples. They are dominated by the number of clusters, the individual uncertainties being much smaller than the intrinsic dispersion. 

The top left panel shows that both samples contain a majority of disturbed (\w$>0.01$) objects.  The shape of the distributions differs, the \highz\ sample having a prominent peak of  objects around \w$\sim0.02$. This is reflected in the cumulative distribution, where there is an excess of \highz\ objects at $\mathrm{log}_{10} \langle w \rangle \sim -1.4$. 
However, the fraction of disturbed objects in  the \esz\ sample ($64 \pm 5\%) $ is nearly identical to that in  the \highz\ sample ($67 \pm 10 \%)$, and is consistent within the uncertainties. We investigated if the two samples are representative of the same population by performing the  Kolmogorov-Smirnov  (KS) test.  We determine the p-value of the null hypothesis, that is the two samples are drawn from the same distribution. The $p$-value is  $50\%$,  indicating no significant difference between the \esz\ and \highz\ samples. 

We obtained similar results studying the distribution of  CC clusters using the \csb\ parameter, as shown in the central panels of \figiac{fig:morpho_histos}.  The fraction of CC clusters is low.  The \highz\ sample has a slightly lower fraction of CC objects $(10 \pm 5\%)$ as compared to the \esz\ sample ($15 \pm 4\%)$, but the difference is not significant. Consistently, the KS test yields a high $p$-value of $47\%$. 

Interestingly, the $M$ distributions shown in the right panels of \figiac{fig:morpho_histos} suggest some evolution. While the distributions  have qualitatively similar shapes,  the \highz\ sample has a peak which is clearly shifted towards disturbed  objects.  Furthermore, the fraction of relaxed  clusters in the \esz\ sample, $53 \pm 5\%$, is $50\%$ higher than that of the \highz\ sample, $34 \pm 8\%$, a $2\,\sigma$ effect. This evolution can be seen  in the cumulative distribution as a systematic over-abundance of disturbed  objects in the \highz\ sample as compared to the \subesz\ sample. The KS test yields a smaller  $p$-value of $25\%$, but this is not small enough to reject the null-hypothesis. 

%

\subsection{Mass dependence and redshift evolution}\label{sec:morpho_evo}

For each morphological parameter, we further quantified the relation with mass and redshift by computing the Spearman's rank (SR) coefficient $\rho$ and the null hypothesis  $p$-value, the probability that the observed coefficient is obtained by chance if the two parameters are completely independent.  We also considered the sum square difference  of ranks, $D$,  and the number of standard deviations by which $D$ deviates from its null-hypothesis expected value, $\sigma_\mathrm{D}$. As in the previous section,  we performed  $1000$ bootstrap resamples to estimate these values and their $68\%$ errors.  The top and bottom panels of \figiac{fig:morpho_evo} show each parameter as a function of  redshift and mass, respectively.  The centroid shift, \csb,\ and $M$  parameters are shown in the left, middle, and right panels, respectively, with the SR coefficient and corresponding $p$-value  indicated in the top left of each plot. 

The only parameter for which there is a correlation with $z$ is the combined $M$ parameter, for which  $\rho=-0.2 \pm 0.1$. The correlation is not very significant, with a null-hypothesis $p$-value of $0.01\pm0.07$,  and a standard deviation on the null hypothesis of  $\sigma_{\mathrm D}=2.5\pm1$. The Kendall test gives consistent results. This weak correlation of  $M$ with $z$ comes from the amplification of the positive (but not significant) trend in \w\ versus $z$, while there is no correlation between \csb\ and redshift.  

In summary, consistent with the trend observed in Sect.~\ref{sec:chara_morpho} above, there is weak evidence that  clusters at higher redshift are slightly more disturbed. On the other hand, there is no evidence for any trend with mass, the SR coefficient for all parameters being consistent with zero and the corresponding $p$-value  in the range $20\%$--$50\%$.  This  is in agreement with the mass independence of the dynamical state found  by \citet{bohringer2010} and  \citet{lov17} for local clusters.  We must note however that the mass range is limited in the present sample.

\section{Total mass profile shape}\label{sec:sparsity}

\begin{table*}
\caption{{\footnotesize Mean values of the sparsity and their uncertainties computed in the bins shown in \figiac{fig:spar_vs_morpho} as a function of $z$, $\langle w \rangle$, \csb, and $M$.  The means are computed in  logarithmic space and take  into account both statistical errors and the intrinsic dispersion, estimated  iteratively. The  intrinsic dispersion  $\sigma_\mathrm{log,int}$ in dex are given in the table.  The values between parentheses are the sparsities  computed within the same bins excluding the outliers.}}\label{tab:spar_values}
\begin{center}
\resizebox{1.\textwidth}{!} {
\begin{tabular}{|ccc|ccc|ccc|ccc|}
\hline        
                   z    &               Mean    & $\sigma_\mathrm{log,int}$&    $\langle w \rangle [10^{-2}]$  & Mean & $\sigma_\mathrm{log,int}$  & \csb        & Mean  & $\sigma_\mathrm{log,int}$& $M$ & Mean&$\sigma_\mathrm{log,int}$ \\
\hline
\hline
 $[ 0.05, 0.22]$ & $  2.32\pm 0.12$ &  $ 0.11\pm 0.04$ & $[ 0.10, 0.80]$ & $  2.26\pm 0.08$ & $ 0.05\pm 0.03$ & $[ 0.08, 0.20]$ & $  3.94\pm 0.51$ & $ 0.28\pm 0.09$ & $[ -6.00, -0.60]$ & $   3.53\pm  0.39$ & $  0.25\pm  0.10$ \\
 &  $( 2.06\pm 0.06)$ & & &  $( 2.06\pm 0.05)$ & & &  $( 2.78\pm 0.17)$ & & &  $(  2.71\pm  0.15)$ & \\
 $[ 0.22, 0.54]$ & $  3.11\pm 0.33$ & $ 0.23\pm 0.12$] & $[ 0.80, 1.90]$ & $  2.77\pm 0.23$ & $ 0.17\pm 0.04$ & $[ 0.20, 0.28]$ & $  2.28\pm 0.11$ & $ 0.08\pm 0.02$ & $[ -0.60,  0.50]$ & $   2.55\pm  0.17$ & $  0.13\pm  0.03$ \\
 &  $( 2.52\pm 0.13)$ & & & $( 2.24\pm 0.10)$ & & & $( 2.06\pm 0.07)$ & & & $(  2.11\pm  0.09)$  &\\
 $[ 0.54, 1.20]$ & $  2.73\pm 0.19$ & $ 0.13\pm 0.05$ & $[ 1.90,10.00]$ & $  3.12\pm 0.32$ & $ 0.23\pm 0.11$ & $[ 0.28, 0.60]$ & $  2.33\pm 0.09$ & $ 0.07\pm 0.03$ & $[  0.50,  3.00]$ & $   2.16\pm  0.05$ & $  0.02\pm  0.02$ \\
 &  $( 2.34\pm 0.09)$ & & & $( 2.51\pm 0.12)$ & & & $( 2.10\pm 0.05)$ &  & & $(  2.04\pm  0.05)$ &\\

\hline
\end{tabular}
}
\end{center}
\end{table*}
\begin{figure*}[!t]
\begin{center}
\includegraphics[width=0.8\textwidth]{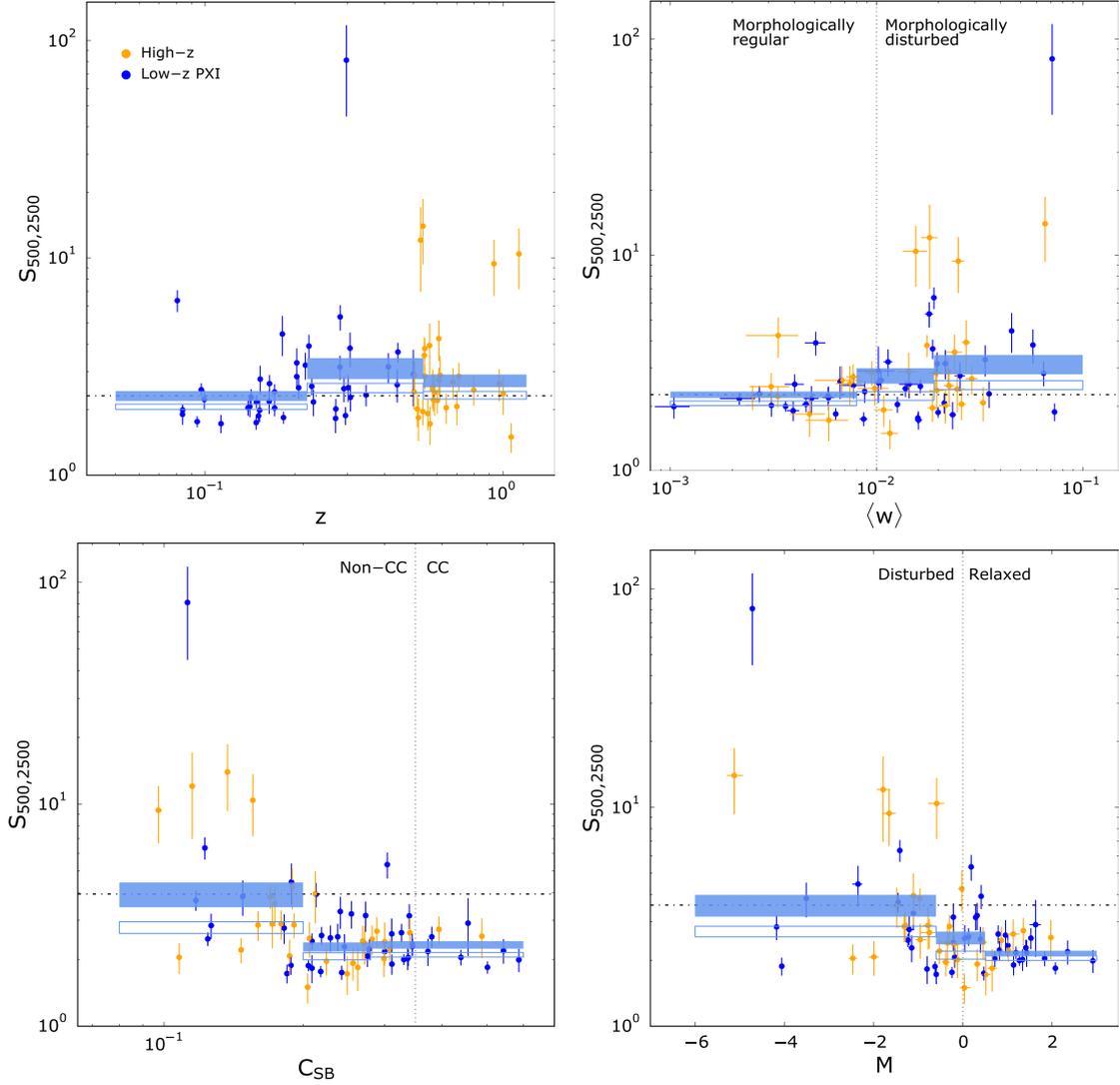} 
\end{center}
\caption{\footnotesize{ \textit{Top left panel:} Sparsity of the \subesz\ and \highz\ samples as a function of redshift. The filled blue rectangles represent the mean sparsity weighted by the statistical errors and the intrinsic dispersion, estimated iteratively (see text), and its $1\sigma$ uncertainty. This is computed in three bins, defined to have approximately   the same number of objects (see Table~\ref{tab:spar_values}). The open blue rectangles represent the same quantity computed removing the outliers. We show as reference the median value of the first bin with the black dotted line. \textit{Top right panel:} Same  but with the sparsity as a function of  $\langle w \rangle$.  \textit{Bottom left panel:} Same but with  sparsity as a function of the \csb.  \textit{Bottom right panel:} Same but with  the sparsity as a function of $M$. The sparsity, i.e. the shape of the total mass profile, varies significantly with the dynamical state indicators. More disturbed clusters have higher sparsity i.e. they are less concentrated. The dependence on redshift is smaller than the dependence on the dynamical state.
}}
\label{fig:spar_vs_morpho}
\end{figure*}

\subsection{Radial mass profiles}

The individual  scaled HE integrated total mass profiles  are shown in \figiac{fig:he_profiles}.   In the left and central panels, the profiles are colour coded according to their morphological state according to the $M$ parameter: in blue for the relaxed ($M>0$) and in red for the disturbed  ($M<0$) clusters. 

There is a clear difference  between the two populations in the $[0.01-0.5]\Rvyx$ range. As compared to the relaxed clusters, the disturbed objects have a shallower mass distribution  on average (i.e. there is less mass within the central region), and these profiles show a larger dispersion.
This effect is even more evident in the central panel of \figiac{fig:he_profiles}, where the mass profiles are plotted as a function of the total density contrast,  $\Delta$. 
 As $\Delta$ is proportional to the mean total density within a sphere of radius $R_{\Delta}$ (Eq.~\ref{eq:mdelta}),  it decreases with radius, more or less rapidly depending on the steepness of  the density profile.  For density profiles that are very peaked towards the centre, $\Delta$ decreases rapidly,  or equivalently  $M\,(<R_\Delta)$ slowly increases with decreasing $\Delta$. On the other hand, a flat profile within the core would correspond to a constant mean density, thus quasi--constant $\Delta$, and a very steep variation of $M\,(<R_\Delta)$ with $\Delta$ in the central region, with a maximum value of $\Delta$ corresponding  to  that of the core.  This is what is observed for the relaxed and disturbed clusters, respectively. In summary, the dynamical state of the cluster clearly has a very strong impact on the shape of the total mass profile.

The right panel of \figiac{fig:he_profiles} shows the $68\%$ dispersion envelopes of the \subesz\ and \highz\ samples with light and dark green, respectively. The black solid and dotted lines represent the median profiles for the \subesz\ and \highz\ samples, respectively. The envelopes of the two samples are  consistent. However, the median HE mass profile of the \highz\ sample is slightly lower and shallower  than that of  the \subesz\ sample, being lower by $6.5\%$ and $16\%$ at $0.9\Rvyx$ and 0.3$\Rvyx$, respectively. The HE radial  mass profiles thus show a hint of evolution, but this is not statistically significant considering the dispersion of the profiles. Interestingly, the dispersions of the two samples are similar. We recall that the samples contain a similar number of objects, the \subesz\ and \highz\ having 42 and 33 objects, respectively, and the data quality ensures that HE mass profiles are computed for each object. In view of our finding above that the morphology evolution is negligible, the absence of  evolution of  the HE mass profile dispersion  is a natural consequence of their shape being driven by the dynamical state. 

\subsection{The shape of the mass  profiles}
\label{sec:sparobs}

We  further quantified  the evolution and the impact of dynamical state on the shape of the profile  using the sparsity, \S.  We chose $\Delta_2=2500$, which  is large enough to encompass the dependence of the mass profile shape as a function of the dynamical state and is reached by all haloes, as shown in the middle  panel of Fig.~\ref{fig:he_profiles}.

Figure \ref{fig:spar_vs_morpho} shows the sparsity as a function of the redshift in the top-left panel, and the three morphological parameters, \w,  \csb,\ and $M$   in the top-right, bottom-left, and bottom-right panels,  respectively. The statistical errors on the sparsity are not negligible as compared to the intrinsic scatter, and the simple correlation tests used in the previous section cannot be applied. We therefore computed the mean sparsity in bins. We defined three bins in redshift  and for each parameter, the width of each bin being defined so as to have roughly the same number of objects in each bin. The logarithmic mean of the sparsity was computed in each bin, weighting each value by the quadratic sum of the statistical  error  and the intrinsic scatter. This scatter and  the  weighted mean  were  estimated simultaneously by iteration.  The  mean and intrinsic scatter, together with their  $68\%$ errors,  were computed using 1000 bootstrap resamples.  The results are reported in Table \ref{tab:spar_values} and as filled blue rectangles in each panel of Fig~\ref{fig:spar_vs_morpho}. As there is a large scatter with the presence of strong outliers, we also computed the mean within the same bins excluding the $>3 \sigma$ outliers. The results  are shown with the open blue rectangles. 

 There seems to be a slight although not very significant evolution of the sparsity with $z$. Higher-redshift ($z>0.54$) clusters have a slightly larger value of \S\ by $18\%$,  that is these profiles are less concentrated, which is not consistent with the  low-redshift ($z<0.2$) clusters at  $\sim 1.8\sigma$. We found the same behaviour when excluding the outliers. 
 
 There is  a much stronger and significant variation  of the sparsity with dynamical state:  morphologically disturbed (high \w),  non-CC (low \csb), and disturbed (low $M$) clusters have larger sparsity. For all parameters considered, the sparsity of the first and third bins is not consistent at more than $3\sigma$, and the difference between the sparsities of these bins is of the order of $\sim 50\%$.   At the same time, the intrinsic scatter increases significantly, reaching $\sim 0.2$ dex for the  most disturbed or least concentrated objects.  For example,  the sparsity increases by $64\%$ at a significance level of $3.4\sigma$ between  $M<-0.6$ and $M>0.5$, that is, between the  one third most relaxed and the one third most disturbed as defined by this parameter. Only an upper limit on the intrinsic scatter can be estimated from the former ($<0.04$ dex), while the intrinsic  scatter for the latter reaches $0.25\pm0.09$ dex.  There are  strong outliers at  high sparsity.  Excluding the outliers yields the same qualitative results, although the variation between bins is weaker.  This suggests  a sparsity  distribution skewed towards high values, with a skewness increasing with departure from  dynamical relaxation. 

\section{Discussion}\label{sec:discussion}
\subsection{Dynamical state}

The first result of our study is that SZ-selected samples are dominated at all redshifts by disturbed and non-CC objects.
Recent observational work on local clusters has converged to similar results  (e.g. \citealt{lov17}, \citealt{rossetti2017}, \citealt{santos2017}, and \citealt{lopes2018}). In particular, these works highlight the higher fraction of disturbed objects or the lower fraction of CC objects in SZ-selected samples as compared to X--ray-selected samples, a fact interpreted to be due to preferential detection of relaxed or more concentrated clusters in X--ray surveys. 

In contrast, there is little consensus in the literature 
 concerning  the evolution of  the dynamical state,  as determined from various morphological parameters. Studies of the evolution up to $z\sim1$ of the centroid shift and/or power ratios of X--ray-selected clusters indicate a larger fraction of disturbed clusters at high $z$  \citep{maughan2008,jeltema2005,weissmann2013}. However,  the latter study is also  consistent with no evolution, and  \citet{nurgaliev2017} did not find any significant  evolution in the redshift range $0.3<z<1$ for the 400d X--ray-selected clusters.  

These somewhat contradictory results may simply be due to selection effects:  X-ray detectability is clearly not independent of cluster morphology. More peaked clusters (usually relaxed)  are not only more luminous at a given mass, but are also easier to  detect at a given X--ray luminosity. Such effects are particularly important in  flux-limited surveys, as shown by \citet{chon2017} in the context of a volume-limited X--ray survey. It is therefore difficult to disentangle selection effects and/or $z$ and mass dependence \citep[see also the discussion in ][]{mantz2015}. 

Sunyaev-Zeldovich detection does not suffer from these limitations, and an SZ-selected sample is expected to be close to mass-limited. The morphological evolution of SPT clusters was studied recently by  \cite{nurgaliev2017} using  their newly introduced  $a_{\mathrm phot}$ parameter. They did not find a significant  difference between  the redshift ranges $[0.3$--$0.6]$ and $[0.6$--$1.2]$. This absence of  significant evolution was also observed by \cite{macdonald2017} up to $z=1.6$, also for an SPT sample.  The comprehensive study of classification criteria for the most relaxed clusters  by \citet[][]{mantz2015} indicates that the fraction of relaxed clusters in the SPT and \planck\ samples is consistent with being constant with redshift. 
  
In the present study, we extend the morphological analysis of the full SZ-selected population of high-mass clusters,   from very local systems $z=0.05$  up to $z=1$, applying a consistent sample construction and analysis strategy over the full $z$ range.  The high quality of our data allows us to investigate core (i.e. \csb) and bulk (i.e. \w) properties at high precision. There are no significant trends either with $z$ or mass in these parameters individually. However, we find  a significant evolution with $z$ of the  $M$ dynamical indicator, which  combines these large-scale and core parameters, with a null-hypothesis  $p$-value of $1\%$.  

It has  been suggested that  the fraction of disturbed clusters should increase with $z$ and mass in a hierarchical formation model \citep[e.g.][]{bohringer2010,macdonald2017}. To explain the observed absence of evolution in the SPT sample, \citet{macdonald2017} proposed a simple model, combining the merger rate from  the simulations of  \citet{fakhouri2010}  and  a fiducial relaxation time of the hot gas equal to the crossing time.  In fact, the link between cluster formation history and morphological state as observed in X-rays, as a function of $z$ and mass,  is very complex. This first depends on relating the individual  mass assembly history  to dynamical state \citep[e.g.]{power2012}, and then the dynamical state to morphological indicators \citep[e.g.][]{cui2017}.  Individual cluster history is never observed directly and has to be translated into the ensemble properties of cluster samples at different $z$ \citep[e.g. see][]{mostoghiu2019}.  A further complication is  the relation between the gas dynamical history and that of the underlying dark matter, and how X--ray morphological observables relate to the gas dynamical state.  To our knowledge the only theoretical prediction of the evolution of observed ICM  morphological  parameters  is that of \citet{jeltema2008}. These latter authors   claim a significant  evolution of the mean \w\ with $z$, although a comparison of their results with our data in Fig.~\ref{fig:morpho_evo} shows that any such evolution is  mild, is much smaller than the dispersion, and is fully in agreement with our results.

\subsection{Total mass  profile}
  
 Extending the pilot study of \citet{bartalucci18} with a fully SZ-selected sample, the overall picture emerging from the present work  is that the shape of the dark matter profiles is affected both by evolution and by dynamical state (Table \ref{tab:spar_values}). The evolution effect is mild, increasing the sparsity of objects by  $\sim15\%$ from  $z \sim 0.1$  to  $z \sim 0.8$. In contrast, the  $M$ dependence is much stronger,  with a sparsity increasing  by $\sim 60\%$ with decreasing  $M$, that is from  the  most relaxed to the most disturbed objects.  This variation of  profile shape with dynamical state  is  likely  a fundamental property, rather than a secondary consequence of the mutual variation of the $S_{500,2500}$ and $M$  parameters with $z$, which are both  less significant. A multi-component analysis requiring a larger sample  is needed to firmly assess this point. 
 
  An obvious question is whether the observed dependence of the sparsity on dynamical state is an artefact of systematic error in the X-ray mass estimate. As discussed in detail by \citet{corasaniti2018} the sparsity derived from HE mass estimates is essentially bias-free (less than $5\%$). As it is a mass ratio, the sparsity  is only sensitive to the radial dependence of any bias, which is usually small between the density contrasts under consideration. The mean $S_{500,2500}$ bias from the  HE mass estimate for example is of $\sim3.2\%$  from  the simulations  of  \citet{biffi2016}.
Generally, although the exact radial dependence of the HE bias will differ from object to object, we expect it to  increase towards larger radii (smaller $\Delta$) meaning that sparsity \S\ measured from HE profiles would be { biased low} as compared to the true value, especially for the most disturbed objects. This effect is the opposite to the observed  { increase} of sparsity for increasingly perturbed systems.  As detailed in Sect.~\ref{sec:spardef} we are not directly using the HE mass at $\Delta=500$ but its proxy, $\Mvyx$. For the clusters for which no extrapolation is needed, the differences between these measures are the order of $10\%$, with  no systematic trend with dynamical state. Thus, the measured sparsity may be slightly  lower than if we had used the HE mass, but the clear trend of the sparsity with dynamical state would not be changed.  

The trend of sparsity with  dynamical state indicates that morphologically disturbed objects are less concentrated than relaxed objects. There is also  evidence  for increased  scatter.  This is qualitatively in agreement with the difference in the $c$--$M$ relations for relaxed versus disturbed objects seen in numerical simulations \citep[e.g.][Fig.~1 and 4]{bhattacharya2013}.  In \citet{lebrun2018} we performed a preliminary investigation using numerical simulations tailored to cover the high-mass high-z range considered here. 
In these simulations, the sparsity of the 25 most massive clusters at all redshifts  shows a correlation with the DM dynamical indicator $\Delta$r\footnote{$\Delta$r is defined as the distance between the centre of mass and the centre of the shrinking sphere \citep{lebrun2018}.} with a $p$-value of $[0.5-2]\%$ (their Fig. 3), indicating that sparser clusters are less regular. We will revisit the link between dynamical state and DM sparsity for the full simulated sample in a forthcoming  paper.

\section{Conclusions}\label{sec:conclusions}

We present new \xmm\ observations of a \planck\ SZ-selected sample of 28 massive clusters in the redshift range $z=[0.5$--$0.9]$. These were combined  with the sample of \citet{bartalucci18} at $0.9<z<1.1$ for a total \highz\ sample of 33 objects at masses $\Mvysz > 5 \times 10^{14} M_{\odot}$. 
We characterised the dynamical state with  the centroid shift \w, the concentration  \csb, and the combination of the two parameters, $M$,  which simultaneously probes  the large-scale and  core morphology. The shape of the total mass profile, derived from the hydrostatic equilibrium equation,  was quantified using the sparsity, the ratio of $\Mv$ to $M_{2500}$, that is the masses at density contrast 500 and 2500, respectively.  This parameter, introduced by \citet{balmes2014}  offers a non-parametric measurement of the shape which is thought to be relatively insensitive to HE bias \citep{corasaniti2018} . 

We first combined  the \highz\  morphology measurements with those  of  the ESZ clusters at $z<0.5$ in  \citet{lov17}, for a total sample of  151 objects. In this study:
\begin{itemize}
\item  We confirmed that  SZ-selected samples, thought to best reflect the underlying cluster population, are dominated by disturbed ($\sim65 \%$) and non-CC ($ \sim 80\%$) objects, at all redshifts. 
\item There is no significant evolution or mass dependence of the fraction of cool core  or of  the centroid shift parameter.  The only parameter for which there is a significant correlation with $z$ is the combined $M$ parameter, for which  $\rho=-0.2 \pm 0.1$ and a null-hypothesis $p$-value of $0.01$.
\end{itemize}

We then combined the  mass measurements obtained for our new data with those from a subsample of 42 ESZ objects  with spatially resolved ICM profiles presented in \cite{planck_pressure}, and which we confirmed  is representative of the full \esz\  sample. The total sample of 75 objects covers the redshift range $0.08<z<1.1$ and mass range $[5-20] \times 10^{14} M_{\odot}$. We made the following findings.  
\begin{itemize}
\item The median scaled mass profile differs by less than $6.5\%$ and $16\%$ at $0.9\Rvyx$ and 0.3$\Rvyx$, respectively, between the \subesz\  and \highz\  samples, with no  difference in the dispersion. The evolution of the sparsity with $z$ is mild: it  increases by only  $18\%$ between $z<0.2$ and $z>0.55$, an effect significant at less than  $2\sigma$. 
\item When expressed in terms of  a scaled mass profile, there is a clear difference  between  relaxed and disturbed objects. The latter  have a less concentrated mass distribution on average, and their scaled profiles show a much larger dispersion.
\item  Consequently  there is a clear dependence of the sparsity  on  the dynamical state. When expressed as a function of the $M$ parameter,  the sparsity   increases by  $\sim 60\%$ from the one third most relaxed  to the one third most disturbed objects, an effect significant at  more than  $3\sigma$ level. We discussed the fact that the HE bias will not
significantly change this result. 
\end{itemize}

The main result of this work is that the radial mass distribution  is chiefly governed by  the dynamical state of the cluster and only mildly dependent  on redshift. 
This has  important consequences:
\begin{itemize}
\item A coherent sample selection at all $z$ is key. For instance, one cannot compare the $c$--$M$ relation calibrated at low $z$ on the most relaxed  X-ray selected clusters to that of a SZ-selected sample at high  $z$.  Ideally, one should consider complete samples,   representative of the full  true underlying cluster population; for example, a mass-selected sample. This is even more critical when comparing theory and observation, in view of the difficulty of defining coherent dynamical indicators between the two.  
\item To test theoretical predictions, it is insufficient to simply compare median or stacked properties at each $z$. The dispersion is a critical quantity, as is the profile distribution, in view of likely departure from log-normality. This requires the measurement of  individual profiles. X--ray observations currently provide the best way to obtain such profiles at high statistical precision.
\item In view of this, observational and theoretical  efforts to understand the HE bias and its radial dependence  are all the more important.   
\end{itemize}

In a forthcoming paper, we will extend our study of the dependence of the sparsity on the dynamical state using  an extension of the dark matter simulations presented in \citet{lebrun2018} to a larger sample of  clusters at $\Mv>5\times10^{14} M_{\odot}$. On a longer timescale, the link between the true dark matter  distribution and its dynamical state and the X-ray observables will need to be better understood. This includes the link between ICM morphological proxies  and the true underlying dynamical state, and the potential critical issue of the radial variation of the HE bias.  We will address these issues with dedicated simulations.

On the observational side, we will investigate the HE bias by comparing our results to weak lensing mass measurements for a subsample of the present data set. The observed lack of significant evolution needs to be tested with a larger sample, particularly at $z>0.7$ where the present sample is limited, with data of the same or better quality. The fundamental link uncovered by the present paper between the mass profile and dynamical state will  also need to be consolidated with better low$-z$ data.  Our current low$-z$ sample relies on archival data of uneven quality and is not a complete sample.   The building of a new local reference SZ-selected sample, with high-quality ICM thermodynamic and HE mass profiles, will be one of the main outcomes of the AO17 \xmm\ heritage program `Witnessing the culmination of structure formation in the Universe', and will provide the necessary inputs.

\begin{acknowledgements} 
The authors would like to thank Barbara Sartoris for helpful comments and suggestions. The results reported in this article are based on data obtained with \xmm, an ESA science mission with instruments and contributions directly funded by ESA Member States and NASA. This work was supported by CNES. The research leading to these results has received funding from the European Research  Council  under  the  European  Union's  Seventh  Framework
Programme (FP72007-2013) ERC grant agreement no 340519. LL acknowledges support from NASA through contracts 80NSSCK0582 and 80NSSC19K0116.

 \end{acknowledgements}

\bibliographystyle{aa}
\bibliography{lib_articoli}
\appendix
\section{\esz\ versus \subesz\ characterisation}\label{sec:ll_vs_jd}
\begin{figure*}[!ht]
\begin{center}
\includegraphics[width=.8\textwidth]{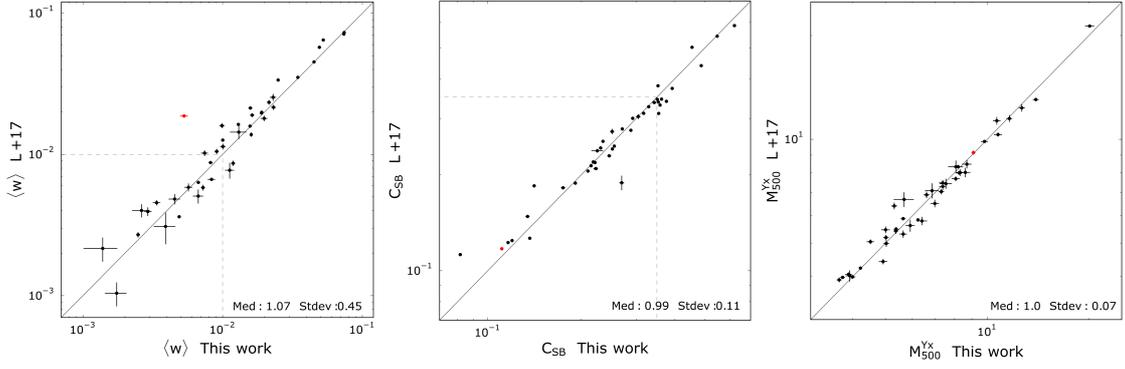} 
\end{center}
\caption{\footnotesize{\textit{Left panel:} Comparison between the centroid shifts computed in this work and published in \cite{lov17} on the \textit{x} and \textit{y} axis, respectively. The red points highlight the  outlier. The dotted lines indicate the threshold used to discriminate between disturbed and relaxed clusters. The solid line is the identity relation. The median and standard deviation were computed weighting by the errors and excluding the two outliers. \textit{Central panel:} Same as the left panel except that the comparison is done for the \csb. The dotted lines indicate the threshold used to discriminate between CC and non-CC clusters. \textit{Right panel:} Same as the left panel except that the comparison is done for the $\Mvyx$.}}
\label{fig:jd_vs_ll}
\end{figure*}
In this work we used the results of \cite{lov17} to characterise the morphological properties of the \esz\ sample, using it as anchor for the local universe properties. For this reason, it is mandatory that the morphological parameters we derived for the \highz\ sample are coherent with the values computed from \cite{lov17}. We derived the centroid shift, \csb, and the $\Mvyx$ for the \subesz\ objects we have in common with \cite{lov17}. The comparison between these values are shown in \figiac{fig:jd_vs_ll}, denoting our and \cite{lov17} values with "This work" and "L+17" labels, respectively.
The measurements of the centroid shifts shown in the left panel are in good agreement, with a ratio of $1.07$ and a standard deviation of $0.45$. The strongest outlier is MACSJ2243.3-0935, shown with a red point. The difference is probably caused by the different choice of the X-ray peak (distant by $\sim7\arcsec$) amplified by the high ellipticity of this object. The concentration parameter \csb\ and the masses, shown in the central and right panels, respectively, are in excellent agreement. The median ratio for both quantities is excellent and the dispersion is remarkably small, the two analysis being performed with different pipelines. 
This comparison shows that the measurements that require larger samples for statistical reasons, such as the centroid shift, are more sensitive to analysis parameters such as the choice of the centre or the exclusion of point sources; integrated quantities such as the \csb\ are more robust.

\section{Representativeness of the \subesz}\label{sec:esz_repr}
We derived the spatially resolved HE mass profiles for a subsample of 42 clusters of the \esz\ sample, namely the \subesz. This subsample covers the same \esz\ redshift range and comprises clusters with $\Mvyx \ge 4\times 10^{14} M_{\odot}$. We investigated if this subsample is representative of the \esz\ population in terms of morphological status. 

\begin{figure*}[!ht]
\begin{center}
\includegraphics[width=.8\textwidth]{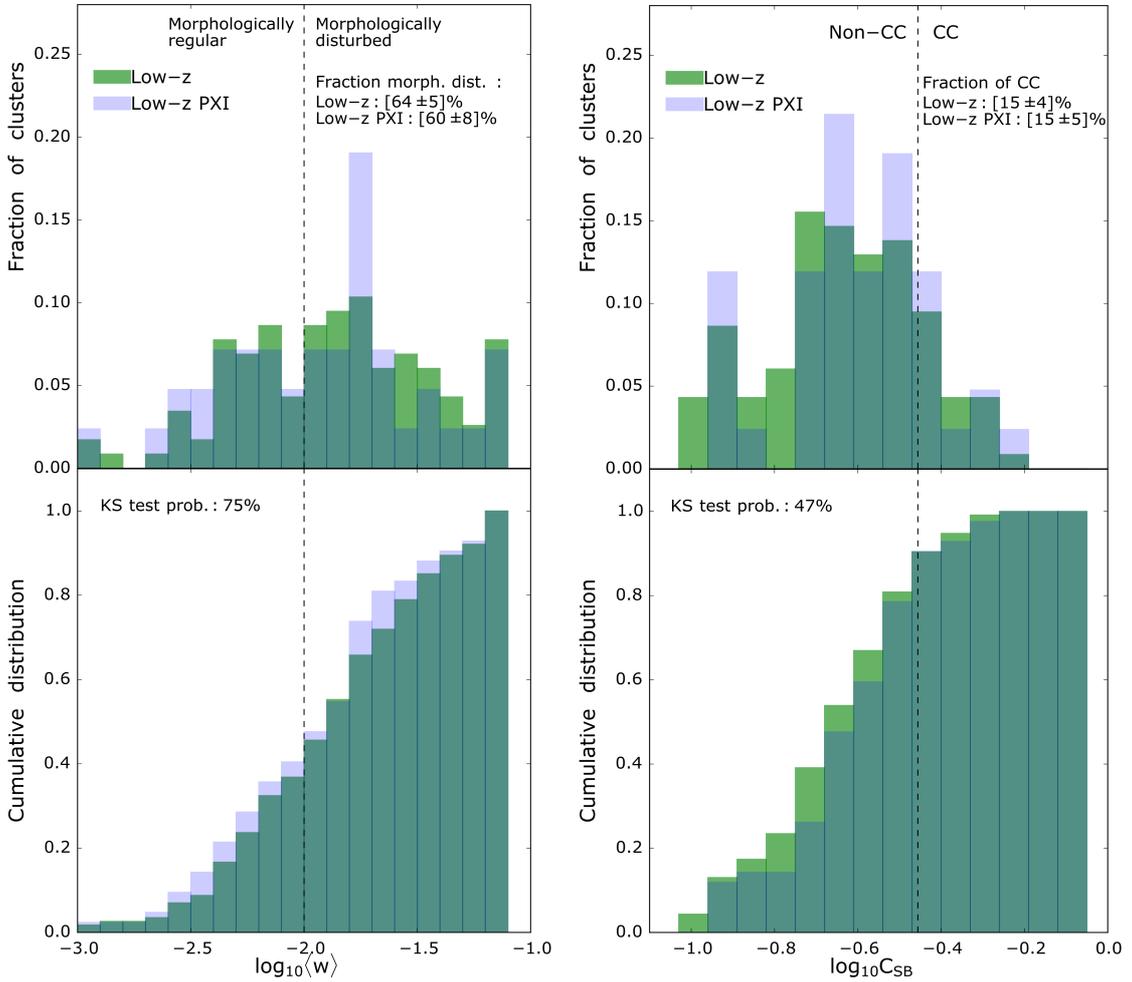} 
\end{center}
\caption{\footnotesize{ \textit{Left top and bottom panels:} Normalised histogram and cumulative distribution of the centroid shift distribution of the \esz\ and \subesz\ samples in green and blue, respectively. We report the probability that the two samples are correlated using the KS test.
\textit{Left top and bottom panels:} Same as the left top and bottom panels, except that we show the normalised and cumulative distribution of the \csb\ in the top and bottom panels, respectively.}}
\label{fig:Represent}
\end{figure*}

The \esz\ and the \subesz\ $\langle w \rangle$ distributions are shown in the left panels of \figiac{fig:Represent} with green and blue polygons, respectively. In particular, the normalised and cumulative distributions are shown in the left and bottom panels, respectively.  The two samples present qualitatively the same distribution and have a similar fraction of disturbed clusters, the \esz\ and the \subesz\ yielding a fraction of disturbed clusters equal to $[65 \pm 2]\%$ and to $[60 \pm 2]\%$, respectively. This result is confirmed by the high value of the KS test probability: $60\%$. We found similar results using the \csb. The cluster distributions as a function of this parameter are shown in the right top and bottom panels of \figiac{fig:Represent}. 
The two samples have the same fraction of CC objects (15\%), and the KS test probability value of $40\%$ confirms that they are representative of the same population. 
        
\clearpage
\vspace{10cm}
\begin{sidewaystable*}[!ht]
\caption{ {\footnotesize Observational and derived dynamical and morphological properties of the \subesz\ and \highz\  samples used in this work.  The horizontal line divides the \subesz\  and the \highz\ sample. 
The $N_{H}$ column reports the neutral hydrogen column density integrated along the line of sight determined from the LAB survey \citep{kalberla2005}.
The Exp. column reports the exposure times after the flare cleaning procedures. The \w and \csb values  of the \subesz\ clusters  were  published by  \citet{lov17}, as well as the properties of the 5 clusters at $z>0.9$ by \cite{bartalucci18} . They are listed for completeness. The \csb\ values of the \highz\ sample have been $k$-corrected as described in Sect. \ref{sec:csb}.}}\label{tab:500_prop}
\begin{center}
\resizebox{0.9\textwidth}{!} {
\begin{tabular}{llllcccccccccccccccccc}
\hline        
\hline
\planck\ name & SPT name & ACT name & Alt. name & z & RA & DEC & $N_{\mathrm H}$  & Exp.  & Obs. Id & $\Rvyx$ & $\Mvyx$               & $M_{2500}^\mathrm{HE}$ & S &  $\langle w \rangle$         & $C_{SB}$ & $kC_{SB}$ & M \\
& &   &   &  &  & &  &  MOS1-2, PN & &    &  &  &   &   &        &                 & \\
 & &   &   &  & [J2000] & [J2000] & [$10^{20} $cm$^{-3}$] &  [ks] & &  [kpc]  & [$10^{14} M_{\odot}$] & [$10^{14} M_{\odot}$]  &   &  $[10^{-2}]$  &   [$10^{-2}$]       &           [$10^{-2}$]        & \\
\hline

PSZ2 G093.92+34.92 &  & &A2255 & $ 0.081 $ & $ 258.2029 $ & $  64.0636 $ & $  2.50 $ & $            6 $, $            4 $ & 0112260801& $         1176_{-          12}^{+          16}$ & $  5.00_{- 0.15}^{+ 0.21}$ & $  0.79_{- 0.70}^{+ 0.87}$ & $  6.36_{- 0.73}^{+ 0.74}$ & $  1.90_{- 0.06}^{+ 0.06}$ & $ 12.26_{- 0.00}^{+ 0.00}$ &    & $ -1.41_{- 0.03}^{+ 0.05}$\\
PSZ2 G306.77+58.61 &  & &A1651 & $ 0.084 $ & $ 194.8438 $ & $  -4.1983 $ & $  1.81 $ & $            7 $, $            4 $ & 0203020101& $         1135_{-          10}^{+          10}$ & $  4.51_{- 0.11}^{+ 0.12}$ & $  2.37_{- 2.13}^{+ 2.61}$ & $  1.90_{- 0.20}^{+ 0.20}$ & $  0.39_{- 0.03}^{+ 0.03}$ & $ 31.14_{- 0.12}^{+ 0.12}$ &    & $  1.14_{- 0.02}^{+ 0.02}$\\
PSZ2 G306.66+61.06 &  & &A1650 & $ 0.084 $ & $ 194.6734 $ & $  -1.7622 $ & $  0.72 $ & $           34 $, $           28 $ & 0093200101& $         1110_{-           4}^{+           4}$ & $  4.21_{- 0.05}^{+ 0.05}$ & $  2.11_{- 2.00}^{+ 2.22}$ & $  2.00_{- 0.11}^{+ 0.11}$ & $  0.36_{- 0.01}^{+ 0.01}$ & $ 33.05_{- 0.03}^{+ 0.03}$ &    & $  1.27_{- 0.00}^{+ 0.00}$\\
PSZ2 G321.98-47.96 & SPT-CLJ2249-6426 & &A3921 & $ 0.094 $ & $ 342.4917 $ & $ -64.4294 $ & $  1.61 $ & $           30 $, $           23 $ & 0112240101& $         1062_{-           6}^{+           6}$ & $  3.73_{- 0.06}^{+ 0.06}$ & $  2.12_{- 2.00}^{+ 2.24}$ & $  1.76_{- 0.10}^{+ 0.10}$ & $  1.59_{- 0.04}^{+ 0.04}$ & $ 21.84_{- 0.00}^{+ 0.00}$ &    & $ -0.24_{- 0.02}^{+ 0.02}$\\
PSZ2 G336.60-55.43 &  & &A3911 & $ 0.097 $ & $ 341.5711 $ & $ -52.7261 $ & $  1.50 $ & $           23 $, $           10 $ & 0149670301& $         1085_{-           7}^{+           7}$ & $  3.99_{- 0.08}^{+ 0.08}$ & $  1.61_{- 1.50}^{+ 1.72}$ & $  2.47_{- 0.18}^{+ 0.17}$ & $  1.63_{- 0.04}^{+ 0.04}$ & $ 12.43_{- 0.01}^{+ 0.01}$ &    & $ -1.23_{- 0.02}^{+ 0.02}$\\
PSZ2 G332.23-46.37 & SPT-CLJ2201-5956 & &A3827 & $ 0.099 $ & $ 330.4720 $ & $ -59.9454 $ & $  2.09 $ & $           21 $, $           11 $ & 0149670101& $         1217_{-           7}^{+           7}$ & $  5.64_{- 0.09}^{+ 0.09}$ & $  2.53_{- 2.28}^{+ 2.78}$ & $  2.23_{- 0.22}^{+ 0.22}$ & $  0.58_{- 0.03}^{+ 0.03}$ & $ 27.83_{- 0.05}^{+ 0.05}$ &    & $  0.82_{- 0.01}^{+ 0.02}$\\
PSZ2 G053.53+59.52 &  & &A2034 & $ 0.113 $ & $ 227.5528 $ & $  33.5104 $ & $  1.54 $ & $           10 $, $            6 $ & 0303930101& $         1211_{-          10}^{+          10}$ & $  5.64_{- 0.14}^{+ 0.14}$ & $  3.27_{- 2.96}^{+ 3.57}$ & $  1.73_{- 0.17}^{+ 0.17}$ & $  1.60_{- 0.07}^{+ 0.07}$ & $ 18.45_{- 0.01}^{+ 0.01}$ &    & $ -0.60_{- 0.04}^{+ 0.03}$\\
PSZ2 G241.79-24.01 &  & &A3378 & $ 0.139 $ & $  91.4752 $ & $ -35.3023 $ & $  4.02 $ & $           14 $, $           11 $ & 0201901001& $         1061_{-           8}^{+           8}$ & $  3.88_{- 0.09}^{+ 0.09}$ & $  1.90_{- 1.76}^{+ 2.05}$ & $  2.04_{- 0.16}^{+ 0.16}$ & $  0.46_{- 0.02}^{+ 0.02}$ & $ 43.93_{- 0.10}^{+ 0.10}$ &    & $  1.83_{- 0.01}^{+ 0.02}$\\
PSZ2 G002.77-56.16 &  & &A3856 & $ 0.141 $ & $ 334.6677 $ & $ -38.9018 $ & $  1.13 $ & $           14 $, $            4 $ & 0201903001& $         1146_{-          11}^{+          11}$ & $  4.92_{- 0.14}^{+ 0.14}$ & $  2.38_{- 2.13}^{+ 2.62}$ & $  2.07_{- 0.22}^{+ 0.22}$ & $  2.13_{- 0.04}^{+ 0.04}$ & $ 27.56_{- 0.09}^{+ 0.09}$ &    & $ -0.18_{- 0.03}^{+ 0.03}$\\
PSZ2 G226.18+76.79 &  & &A1413 & $ 0.143 $ & $ 178.8250 $ & $  23.4049 $ & $  1.84 $ & $           62 $, $           42 $ & 0502690201& $         1240_{-           4}^{+           4}$ & $  6.24_{- 0.06}^{+ 0.06}$ & $  2.73_{- 2.49}^{+ 2.97}$ & $  2.28_{- 0.20}^{+ 0.20}$ & $  0.27_{- 0.01}^{+ 0.01}$ & $ 34.50_{- 0.06}^{+ 0.06}$ &    & $  1.41_{- 0.01}^{+ 0.01}$\\
PSZ2 G236.92-26.65 &  & &A3364 & $ 0.148 $ & $  86.9081 $ & $ -31.8727 $ & $  2.07 $ & $           22 $, $           14 $ & 0201900901& $         1151_{-           7}^{+           8}$ & $  5.01_{- 0.10}^{+ 0.10}$ & $  2.88_{- 2.67}^{+ 3.08}$ & $  1.74_{- 0.13}^{+ 0.13}$ & $  0.87_{- 0.04}^{+ 0.04}$ & $ 24.25_{- 0.09}^{+ 0.09}$ &    & $  0.47_{- 0.03}^{+ 0.03}$\\
PSZ2 G008.47-56.34 &  & &A3854 & $ 0.149 $ & $ 334.4400 $ & $ -35.7260 $ & $  1.20 $ & $           22 $, $           14 $ & 0201902901& $         1035_{-           6}^{+           8}$ & $  3.64_{- 0.06}^{+ 0.09}$ & $  1.68_{- 1.58}^{+ 1.78}$ & $  2.17_{- 0.14}^{+ 0.14}$ & $  0.22_{- 0.04}^{+ 0.04}$ & $ 30.04_{- 0.10}^{+ 0.10}$ &    & $  1.19_{- 0.03}^{+ 0.03}$\\
PSZ2 G003.93-59.41 &  & &A3888 & $ 0.151 $ & $ 338.6125 $ & $ -37.7360 $ & $  1.32 $ & $           20 $, $           13 $ & 0404910801& $         1304_{-           8}^{+           8}$ & $  7.31_{- 0.14}^{+ 0.14}$ & $  3.90_{- 3.70}^{+ 4.11}$ & $  1.87_{- 0.11}^{+ 0.11}$ & $  1.99_{- 0.03}^{+ 0.03}$ & $ 20.53_{- 0.03}^{+ 0.03}$ &    & $ -0.62_{- 0.01}^{+ 0.02}$\\
PSZ2 G021.10+33.24 &  & &A2204 & $ 0.152 $ & $ 248.1959 $ & $   5.5754 $ & $  6.97 $ & $           14 $, $            8 $ & 0306490201& $         1347_{-          13}^{+          13}$ & $  8.06_{- 0.22}^{+ 0.23}$ & $  4.05_{- 3.58}^{+ 4.53}$ & $  1.99_{- 0.24}^{+ 0.24}$ & $  0.10_{- 0.02}^{+ 0.02}$ & $ 58.66_{- 0.23}^{+ 0.23}$ &    & $  2.91_{- 0.02}^{+ 0.01}$\\
PSZ2 G244.71+32.50 &  & &A0868 & $ 0.153 $ & $ 146.3575 $ & $  -8.6557 $ & $  3.59 $ & $            8 $, $            5 $ & 0017540101& $         1058_{-          10}^{+          13}$ & $  3.91_{- 0.11}^{+ 0.15}$ & $  1.42_{- 1.21}^{+ 1.63}$ & $  2.77_{- 0.42}^{+ 0.42}$ & $  2.54_{- 0.15}^{+ 0.15}$ & $ 18.19_{- 0.01}^{+ 0.01}$ &    & $ -1.20_{- 0.10}^{+ 0.09}$\\
PSZ2 G049.22+30.87 &  & &RXJ1720.1+2638 & $ 0.164 $ & $ 260.0417 $ & $  26.6250 $ & $  5.65 $ & $           14 $, $            8 $ & 0500670401& $         1166_{-           8}^{+           8}$ & $  5.30_{- 0.11}^{+ 0.11}$ & $  2.42_{- 2.12}^{+ 2.73}$ & $  2.19_{- 0.28}^{+ 0.28}$ & $  0.58_{- 0.04}^{+ 0.04}$ & $ 54.33_{- 0.20}^{+ 0.20}$ &    & $  2.35_{- 0.03}^{+ 0.03}$\\
PSZ2 G263.68-22.55 & SPT-CLJ0645-5413 & ACT-CL J0645-5413&A3404 & $ 0.164 $ & $ 101.3712 $ & $ -54.2273 $ & $  5.60 $ & $           11 $, $            7 $ & 0404910401& $         1302_{-          11}^{+          10}$ & $  7.38_{- 0.18}^{+ 0.18}$ & $  2.80_{- 2.54}^{+ 3.06}$ & $  2.64_{- 0.25}^{+ 0.25}$ & $  1.05_{- 0.05}^{+ 0.05}$ & $ 32.66_{- 0.38}^{+ 0.38}$ &    & $  0.79_{- 0.04}^{+ 0.03}$\\
PSZ2 G097.72+38.12 &  & &A2218 & $ 0.171 $ & $ 248.9597 $ & $  66.2125 $ & $  2.60 $ & $           17 $, $           11 $ & 0112980101& $         1169_{-           7}^{+           7}$ & $  5.37_{- 0.10}^{+ 0.10}$ & $  2.23_{- 2.04}^{+ 2.41}$ & $  2.41_{- 0.21}^{+ 0.21}$ & $  1.38_{- 0.05}^{+ 0.05}$ & $ 20.91_{- 0.06}^{+ 0.06}$ &    & $ -0.21_{- 0.03}^{+ 0.03}$\\
PSZ2 G067.17+67.46 &  & &A1914 & $ 0.171 $ & $ 216.5105 $ & $  37.8243 $ & $  1.06 $ & $           15 $, $            7 $ & 0112230201& $         1348_{-          11}^{+          10}$ & $  8.25_{- 0.19}^{+ 0.19}$ & $  4.05_{- 3.72}^{+ 4.38}$ & $  2.04_{- 0.17}^{+ 0.17}$ & $  1.26_{- 0.01}^{+ 0.01}$ & $ 33.78_{- 0.02}^{+ 0.02}$ &    & $  0.72_{- 0.01}^{+ 0.01}$\\
PSZ2 G149.75+34.68 &  & &A0665 & $ 0.182 $ & $ 127.7462 $ & $  65.8398 $ & $  4.24 $ & $            5 $, $            2 $ & 0109890401& $         1334_{-          23}^{+          28}$ & $  8.07_{- 0.41}^{+ 0.53}$ & $  1.81_{- 1.45}^{+ 2.17}$ & $  4.46_{- 0.93}^{+ 0.94}$ & $  4.52_{- 0.09}^{+ 0.09}$ & $ 18.87_{- 0.99}^{+ 0.99}$ &    & $ -2.35_{- 0.11}^{+ 0.10}$\\
PSZ2 G313.33+61.13 &  & &A1689 & $ 0.183 $ & $ 197.8726 $ & $  -1.3417 $ & $  1.52 $ & $           36 $, $           27 $ & 0093030101& $         1339_{-           6}^{+           6}$ & $  8.19_{- 0.11}^{+ 0.11}$ & $  4.45_{- 4.17}^{+ 4.73}$ & $  1.84_{- 0.12}^{+ 0.12}$ & $  0.63_{- 0.01}^{+ 0.01}$ & $ 50.16_{- 0.07}^{+ 0.07}$ &    & $  2.08_{- 0.01}^{+ 0.01}$\\
PSZ2 G195.75-24.32 &  & &A0520 & $ 0.203 $ & $  73.5402 $ & $   2.9212 $ & $  3.74 $ & $           20 $, $            9 $ & 0201510101& $         1336_{-          11}^{+          12}$ & $  8.31_{- 0.21}^{+ 0.22}$ & $  2.92_{- 2.55}^{+ 3.30}$ & $  2.84_{- 0.37}^{+ 0.37}$ & $  6.46_{- 0.05}^{+ 0.05}$ & $ 12.65_{- 0.01}^{+ 0.01}$ &    & $ -4.17_{- 0.04}^{+ 0.02}$\\
PSZ2 G006.76+30.45 &  & &A2163 & $ 0.203 $ & $ 243.9399 $ & $  -6.1491 $ & $ 16.50 $ & $           10 $, $            6 $ & 0112230601& $         1794_{-          19}^{+          18}$ & $ 20.10_{- 0.62}^{+ 0.61}$ & $  6.12_{- 5.12}^{+ 7.11}$ & $  3.29_{- 0.54}^{+ 0.54}$ & $  3.36_{- 0.01}^{+ 0.01}$ & $ 24.08_{- 0.02}^{+ 0.02}$ &    & $ -1.11_{- 0.01}^{+ 0.00}$\\
PSZ2 G182.59+55.83 &  & &A0963 & $ 0.206 $ & $ 154.2653 $ & $  39.0482 $ & $  1.25 $ & $           18 $, $           12 $ & 0084230701& $         1129_{-           7}^{+           7}$ & $  5.02_{- 0.10}^{+ 0.10}$ & $  1.99_{- 1.77}^{+ 2.20}$ & $  2.53_{- 0.28}^{+ 0.28}$ & $  0.40_{- 0.04}^{+ 0.04}$ & $ 37.99_{- 0.05}^{+ 0.05}$ &    & $  1.53_{- 0.02}^{+ 0.03}$\\
PSZ2 G166.09+43.38 &  & &A0773 & $ 0.217 $ & $ 139.4765 $ & $  51.7315 $ & $  1.28 $ & $           13 $, $           14 $ & 0084230601& $         1232_{-           9}^{+           8}$ & $  6.61_{- 0.14}^{+ 0.14}$ & $  2.06_{- 1.77}^{+ 2.35}$ & $  3.20_{- 0.46}^{+ 0.46}$ & $  1.14_{- 0.05}^{+ 0.05}$ & $ 25.45_{- 0.05}^{+ 0.05}$ &    & $  0.32_{- 0.04}^{+ 0.04}$\\
PSZ2 G092.71+73.46 &  & &A1763 & $ 0.223 $ & $ 203.8298 $ & $  41.0001 $ & $  0.94 $ & $           12 $, $            9 $ & 0084230901& $         1276_{-          11}^{+          11}$ & $  7.38_{- 0.19}^{+ 0.19}$ & $  1.88_{- 1.65}^{+ 2.11}$ & $  3.92_{- 0.50}^{+ 0.50}$ & $  0.51_{- 0.06}^{+ 0.06}$ & $ 21.34_{- 0.02}^{+ 0.02}$ &    & $  0.41_{- 0.03}^{+ 0.03}$\\
PSZ2 G072.62+41.46 &  & &A2219 & $ 0.228 $ & $ 250.0837 $ & $  46.7107 $ & $  1.76 $ & $           12 $, $            6 $ & 0605000501& $         1481_{-          10}^{+          14}$ & $ 11.62_{- 0.24}^{+ 0.33}$ & $  4.54_{- 4.13}^{+ 4.94}$ & $  2.56_{- 0.23}^{+ 0.24}$ & $  1.02_{- 0.05}^{+ 0.05}$ & $ 21.93_{- 0.02}^{+ 0.02}$ &    & $  0.13_{- 0.04}^{+ 0.03}$\\
PSZ2 G073.97-27.82 &  & &A2390 & $ 0.231 $ & $ 328.4031 $ & $  17.6949 $ & $  8.66 $ & $           10 $, $            8 $ & 0111270101& $         1438_{-          13}^{+          13}$ & $ 10.67_{- 0.28}^{+ 0.28}$ & $  4.91_{- 4.24}^{+ 5.59}$ & $  2.17_{- 0.31}^{+ 0.31}$ & $  0.48_{- 0.04}^{+ 0.04}$ & $ 37.27_{- 0.14}^{+ 0.14}$ &    & $  1.43_{- 0.03}^{+ 0.02}$\\
PSZ2 G294.68-37.01 &  & &RXCJ0303.8-7752 & $ 0.274 $ & $  45.9366 $ & $ -77.8784 $ & $  8.73 $ & $           11 $, $            8 $ & 0205330101& $         1261_{-          18}^{+          17}$ & $  7.54_{- 0.31}^{+ 0.31}$ & $  4.13_{- 3.56}^{+ 4.70}$ & $  1.82_{- 0.26}^{+ 0.26}$ & $  2.34_{- 0.09}^{+ 0.09}$ & $ 20.90_{- 0.09}^{+ 0.09}$ &    & $ -0.80_{- 0.05}^{+ 0.06}$\\
PSZ2 G241.76-30.88 &  & &RXCJ0532.9-3701 & $ 0.275 $ & $  83.2323 $ & $ -37.0270 $ & $  2.90 $ & $           11 $, $            6 $ & 0042341801& $         1194_{-          14}^{+          14}$ & $  6.41_{- 0.23}^{+ 0.23}$ & $  3.18_{- 2.83}^{+ 3.52}$ & $  2.02_{- 0.23}^{+ 0.23}$ & $  0.31_{- 0.08}^{+ 0.08}$ & $ 33.67_{- 0.00}^{+ 0.00}$ &    & $  1.34_{- 0.05}^{+ 0.06}$\\
PSZ2 G259.98-63.43 & SPT-CLJ0232-4421 & &RXCJ0232.2-4420 & $ 0.284 $ & $  38.0772 $ & $ -44.3464 $ & $  2.49 $ & $           12 $, $            7 $ & 0042340301& $         1225_{-          12}^{+          12}$ & $  6.99_{- 0.20}^{+ 0.21}$ & $  2.22_{- 1.95}^{+ 2.50}$ & $  3.14_{- 0.40}^{+ 0.40}$ & $  1.97_{- 0.04}^{+ 0.04}$ & $ 33.96_{- 0.06}^{+ 0.06}$ &    & $  0.29_{- 0.03}^{+ 0.02}$\\
PSZ2 G244.37-32.15 &  & &RXCJ0528.9-3927 & $ 0.284 $ & $  82.2211 $ & $ -39.4714 $ & $  2.13 $ & $            7 $, $            4 $ & 0042340801& $         1218_{-          20}^{+          17}$ & $  6.86_{- 0.33}^{+ 0.30}$ & $  1.28_{- 1.12}^{+ 1.45}$ & $  5.35_{- 0.73}^{+ 0.71}$ & $  1.80_{- 0.08}^{+ 0.08}$ & $ 30.43_{- 0.55}^{+ 0.55}$ &    & $  0.19_{- 0.05}^{+ 0.07}$\\
PSZ2 G106.87-83.23 &  & &A2813 & $ 0.292 $ & $  10.8519 $ & $ -20.6229 $ & $  1.54 $ & $           11 $, $            5 $ & 0042340201& $         1155_{-          12}^{+          16}$ & $  5.91_{- 0.18}^{+ 0.25}$ & $  2.37_{- 2.06}^{+ 2.68}$ & $  2.50_{- 0.34}^{+ 0.34}$ & $  0.77_{- 0.10}^{+ 0.10}$ & $ 22.92_{- 0.05}^{+ 0.05}$ &    & $  0.40_{- 0.08}^{+ 0.09}$\\
PSZ2 G266.04-21.25 & SPT-CLJ0658-5556 & ACT-CL J0658-5557&1ES0657-558 & $ 0.296 $ & $ 104.6277 $ & $ -55.9434 $ & $  4.17 $ & $           21 $, $           14 $ & 0112980201& $         1534_{-          11}^{+          11}$ & $ 13.92_{- 0.28}^{+ 0.29}$ & $  7.41_{- 6.72}^{+ 8.09}$ & $  1.88_{- 0.18}^{+ 0.18}$ & $  7.30_{- 0.03}^{+ 0.03}$ & $ 18.82_{- 0.08}^{+ 0.08}$ &    & $ -4.06_{- 0.02}^{+ 0.02}$\\
PSZ2 G195.60+44.06 &  & & A0781 & $ 0.298 $ & $ 140.1018 $ & $  30.5028 $ & $  1.94 $ & $           57 $, $           47 $ & 0401170101& $         1116_{-           6}^{+           6}$ & $  5.36_{- 0.09}^{+ 0.09}$ & $  0.07_{- 0.04}^{+ 0.10}$ & $ 81.13_{-36.44}^{+36.44}$ & $  7.11_{- 0.05}^{+ 0.05}$ & $ 11.24_{- 0.01}^{+ 0.01}$ &    & $ -4.72_{- 0.02}^{+ 0.04}$\\
PSZ2 G125.71+53.86 &  & &A1576 & $ 0.302 $ & $ 189.2441 $ & $  63.1871 $ & $  1.68 $ & $            6 $, $            1 $ & 0402250101& $         1135_{-          27}^{+          29}$ & $  5.67_{- 0.39}^{+ 0.44}$ & $  2.24_{- 1.97}^{+ 2.52}$ & $  2.53_{- 0.35}^{+ 0.37}$ & $  1.44_{- 0.15}^{+ 0.15}$ & $ 23.77_{- 0.33}^{+ 0.33}$ &    & $  0.04_{- 0.10}^{+ 0.10}$\\
PSZ2 G008.94-81.22 &  & &A2744 & $ 0.307 $ & $   3.5775 $ & $ -30.3863 $ & $  1.60 $ & $           14 $, $           10 $ & 0042340101& $         1360_{-          11}^{+          11}$ & $  9.81_{- 0.24}^{+ 0.24}$ & $  2.56_{- 2.10}^{+ 3.01}$ & $  3.84_{- 0.70}^{+ 0.69}$ & $  5.74_{- 0.03}^{+ 0.03}$ & $ 14.80_{- 0.05}^{+ 0.05}$ &    & $ -3.51_{- 0.01}^{+ 0.02}$\\
PSZ2 G278.58+39.16 &  & &A1300 & $ 0.308 $ & $ 172.9775 $ & $ -19.9285 $ & $  4.50 $ & $           11 $, $            9 $ & 0042341001& $         1302_{-          15}^{+          16}$ & $  8.61_{- 0.30}^{+ 0.33}$ & $  3.78_{- 3.27}^{+ 4.29}$ & $  2.28_{- 0.32}^{+ 0.32}$ & $  3.51_{- 0.08}^{+ 0.08}$ & $ 24.58_{- 0.12}^{+ 0.12}$ &    & $ -1.15_{- 0.04}^{+ 0.03}$\\
PSZ2 G349.46-59.95 & SPT-CLJ2248-4431 & &AS1063 & $ 0.347 $ & $ 342.1824 $ & $ -44.5305 $ & $  1.84 $ & $           25 $, $           15 $ & 0504630101& $         1458_{-          10}^{+          10}$ & $ 12.66_{- 0.27}^{+ 0.27}$ & $  5.41_{- 4.80}^{+ 6.02}$ & $  2.34_{- 0.27}^{+ 0.27}$ & $  0.88_{- 0.02}^{+ 0.02}$ & $ 34.45_{- 0.17}^{+ 0.17}$ &    & $  1.01_{- 0.02}^{+ 0.02}$\\
PSZ2 G083.29-31.03 &  & &RXCJ2228+2037 & $ 0.412 $ & $ 337.1405 $ & $  20.6204 $ & $  4.26 $ & $           24 $, $           15 $ & 0147890101& $         1255_{-          14}^{+          13}$ & $  8.70_{- 0.28}^{+ 0.28}$ & $  2.76_{- 2.34}^{+ 3.18}$ & $  3.15_{- 0.49}^{+ 0.49}$ & $  2.16_{- 0.09}^{+ 0.09}$ & $ 27.32_{- 0.53}^{+ 0.53}$ &    & $ -0.21_{- 0.05}^{+ 0.06}$\\
PSZ2 G284.41+52.45 &  & &MACSJ1206.2-0848 & $ 0.441 $ & $ 181.5521 $ & $  -8.8002 $ & $  4.35 $ & $           30 $, $           21 $ & 0502430401& $         1332_{-          12}^{+          12}$ & $ 10.76_{- 0.30}^{+ 0.30}$ & $  4.13_{- 3.43}^{+ 4.83}$ & $  2.61_{- 0.45}^{+ 0.45}$ & $  0.67_{- 0.02}^{+ 0.02}$ & $ 31.12_{- 0.09}^{+ 0.09}$ &    & $  0.96_{- 0.01}^{+ 0.01}$\\
PSZ2 G056.93-55.08 &  & &MACSJ2243.3-0935 & $ 0.444 $ & $ 340.8387 $ & $  -9.5947 $ & $  3.11 $ & $          103 $, $           78 $ & 0503490201& $         1258_{-           6}^{+           6}$ & $  9.10_{- 0.12}^{+ 0.12}$ & $  2.47_{- 2.22}^{+ 2.72}$ & $  3.69_{- 0.38}^{+ 0.38}$ & $  1.87_{- 0.04}^{+ 0.04}$ & $ 11.72_{- 0.07}^{+ 0.07}$ &    & $ -1.45_{- 0.03}^{+ 0.03}$\\
\hline
PSZ2 G265.10-59.50 & SPT-CLJ0243-4833 &  & RXCJ0243.6-4834& $ 0.500 $ & $  40.9129 $ & $ -48.5611 $ & $  2.15 $ & $           15,            6 $  & 0672090501, 0723780801 & $         1079_{-          27}^{+          36}$ & $  6.13_{- 0.45}^{+ 0.63}$ & $  2.11_{- 1.52}^{+ 2.69}$ & $  2.91_{- 0.84}^{+ 0.87}$ & $  1.02_{- 0.21}^{+ 0.20}$ & $ 45.52_{- 1.05}^{+ 1.04}$ & $ 44.77_{- 1.03}^{+ 1.03}$ & $    1.63_{- 0.12}^{+ 0.13}$\\
PSZ2 G044.77-51.30 &  &  & MACSJ2214.9-1359& $ 0.503 $ & $ 333.7383 $ & $ -14.0045 $ & $  2.88 $ & $           17,            7 $  & 0693661901 & $         1175_{-          22}^{+          21}$ & $  7.95_{- 0.44}^{+ 0.44}$ & $  3.29_{- 2.75}^{+ 3.81}$ & $  2.42_{- 0.41}^{+ 0.41}$ & $  0.98_{- 0.17}^{+ 0.16}$ & $ 26.93_{- 0.62}^{+ 0.60}$ & $ 26.55_{- 0.61}^{+ 0.59}$ & $    0.45_{- 0.07}^{+ 0.07}$\\
PSZ2 G211.21+38.66 &  &  & MACSJ0911.2+1746& $ 0.505 $ & $ 137.7970 $ & $  17.7760 $ & $  3.28 $ & $           33,           25 $  & 0693662501 & $         1038_{-          14}^{+          13}$ & $  5.49_{- 0.21}^{+ 0.21}$ & $  1.90_{- 1.62}^{+ 2.18}$ & $  2.89_{- 0.44}^{+ 0.44}$ & $  2.23_{- 0.10}^{+ 0.11}$ & $ 17.92_{- 0.46}^{+ 0.44}$ & $ 17.77_{- 0.46}^{+ 0.44}$ & $   -1.29_{- 0.12}^{+ 0.09}$\\
PSZ2 G004.45-19.55 &  &  & & $ 0.516 $ & $ 289.2692 $ & $ -33.5228 $ & $  5.90 $ & $           14,            5 $  & 0656201001 & $         1207_{-          22}^{+          22}$ & $  8.73_{- 0.48}^{+ 0.48}$ & $  4.34_{- 3.61}^{+ 5.06}$ & $  2.01_{- 0.35}^{+ 0.35}$ & $  2.16_{- 0.15}^{+ 0.14}$ & $ 29.93_{- 0.73}^{+ 0.75}$ & $ 29.56_{- 0.72}^{+ 0.74}$ & $   -0.12_{- 0.14}^{+ 0.15}$\\
PSZ2 G110.28-87.48 &  &  & & $ 0.520 $ & $  12.2939 $ & $ -24.6792 $ & $  1.50 $ & $           26,            4 $  & 0693662101, 0723780201 & $          989_{-          25}^{+          30}$ & $  4.83_{- 0.36}^{+ 0.46}$ & $  2.63_{- 2.09}^{+ 3.17}$ & $  1.84_{- 0.40}^{+ 0.41}$ & $  0.47_{- 0.09}^{+ 0.09}$ & $ 26.27_{- 1.35}^{+ 1.11}$ & $ 26.14_{- 1.34}^{+ 1.11}$ & $    0.66_{- 0.10}^{+ 0.08}$\\
PSZ2 G212.44+63.19 &  &  & RMJ105252.4+241530.0& $ 0.529 $ & $ 163.2159 $ & $  24.2584 $ & $  1.89 $ & $           34,           20 $  & 0693660701, 0723780701 & $          937_{-          18}^{+          17}$ & $  4.15_{- 0.23}^{+ 0.23}$ & $  0.34_{- 0.20}^{+ 0.49}$ & $ 12.06_{- 5.10}^{+ 5.06}$ & $  1.81_{- 0.16}^{+ 0.17}$ & $ 11.51_{- 0.63}^{+ 0.60}$ & $ 11.23_{- 0.61}^{+ 0.58}$ & $   -1.78_{- 0.18}^{+ 0.13}$\\
PSZ2 G201.50-27.31 &  &  & MACSJ0454.1-0300& $ 0.538 $ & $  73.5471 $ & $  -3.0162 $ & $  3.92 $ & $           22,           17 $  & 0205670101 & $         1159_{-          15}^{+          15}$ & $  7.94_{- 0.29}^{+ 0.31}$ & $  4.05_{- 3.51}^{+ 4.59}$ & $  1.96_{- 0.27}^{+ 0.27}$ & $  1.87_{- 0.10}^{+ 0.10}$ & $ 22.46_{- 0.39}^{+ 0.41}$ & $ 22.29_{- 0.38}^{+ 0.40}$ & $   -0.38_{- 0.09}^{+ 0.09}$\\
PSZ2 G094.56+51.03 &  &  & WHL J227.050+57.90& $ 0.539 $ & $ 227.0821 $ & $  57.9164 $ & $  1.50 $ & $           26,           20 $  & 0693660101, 0723780501 & $         1064_{-          14}^{+          14}$ & $  6.15_{- 0.24}^{+ 0.25}$ & $  0.44_{- 0.29}^{+ 0.58}$ & $ 13.98_{- 4.67}^{+ 4.66}$ & $  6.57_{- 0.18}^{+ 0.18}$ & $ 13.72_{- 0.41}^{+ 0.42}$ & $ 13.49_{- 0.41}^{+ 0.41}$ & $   -5.12_{- 0.13}^{+ 0.14}$\\
PSZ2 G228.16+75.20 &  &  & MACSJ1149.5+2223& $ 0.544 $ & $ 177.3976 $ & $  22.4011 $ & $  1.92 $ & $           13,            2 $  & 0693661701 & $         1221_{-          27}^{+          27}$ & $  9.36_{- 0.62}^{+ 0.64}$ & $  2.63_{- 2.11}^{+ 3.15}$ & $  3.56_{- 0.74}^{+ 0.74}$ & $  2.39_{- 0.22}^{+ 0.20}$ & $ 17.37_{- 0.89}^{+ 0.87}$ & $ 17.28_{- 0.88}^{+ 0.87}$ & $   -1.48_{- 0.18}^{+ 0.18}$\\
PSZ2 G111.61-45.71 &  &  & CL0016+16& $ 0.546 $ & $   4.6399 $ & $  16.4362 $ & $  3.99 $ & $           33,           24 $  & 0111000101, 0111000201 & $         1214_{-          11}^{+          11}$ & $  9.21_{- 0.24}^{+ 0.24}$ & $  3.22_{- 2.81}^{+ 3.63}$ & $  2.86_{- 0.37}^{+ 0.37}$ & $  0.95_{- 0.07}^{+ 0.08}$ & $ 19.11_{- 0.30}^{+ 0.30}$ & $ 18.98_{- 0.30}^{+ 0.30}$ & $   -0.30_{- 0.05}^{+ 0.03}$\\
PSZ2 G180.25+21.03 &  &  & MACSJ0717.5+3745& $ 0.546 $ & $ 109.3800 $ & $  37.7587 $ & $  6.63 $ & $          156,          116 $  & 0672420101, 0672420201, 0672420301 & $         1356_{-           6}^{+           6}$ & $ 12.84_{- 0.17}^{+ 0.17}$ & $  3.36_{- 2.96}^{+ 3.75}$ & $  3.82_{- 0.45}^{+ 0.45}$ & $  1.76_{- 0.04}^{+ 0.04}$ & $ 16.98_{- 0.11}^{+ 0.11}$ & $ 16.85_{- 0.11}^{+ 0.11}$ & $   -0.96_{- 0.03}^{+ 0.03}$\\
PSZ2 G183.90+42.99 &  &  & WHL J137.713+38.83& $ 0.559 $ & $ 137.7032 $ & $  38.8357 $ & $  1.63 $ & $           14,            8 $  & 0723780101 & $         1173_{-          25}^{+          27}$ & $  8.44_{- 0.53}^{+ 0.60}$ & $  4.39_{- 3.71}^{+ 5.07}$ & $  1.92_{- 0.32}^{+ 0.33}$ & $  1.08_{- 0.11}^{+ 0.11}$ & $ 25.65_{- 0.71}^{+ 0.73}$ & $ 25.48_{- 0.70}^{+ 0.72}$ & $    0.32_{- 0.08}^{+ 0.06}$\\
PSZ2 G155.27-68.42 &  &  & WHL J24.3324-8.477& $ 0.567 $ & $  24.3536 $ & $  -8.4557 $ & $  3.52 $ & $           28,           18 $  & 0693662801, 0700180201 & $         1149_{-          18}^{+          22}$ & $  8.01_{- 0.38}^{+ 0.46}$ & $  2.03_{- 1.50}^{+ 2.56}$ & $  3.95_{- 1.04}^{+ 1.05}$ & $  2.72_{- 0.13}^{+ 0.13}$ & $ 21.26_{- 0.56}^{+ 0.52}$ & $ 21.02_{- 0.56}^{+ 0.51}$ & $   -1.11_{- 0.09}^{+ 0.12}$\\
PSZ2 G046.13+30.72 &  &  & WHL J171705.5+240424& $ 0.569 $ & $ 259.2742 $ & $  24.0737 $ & $  5.18 $ & $           26,            2 $  & 0693661401 & $          843_{-          20}^{+          19}$ & $  3.17_{- 0.22}^{+ 0.22}$ & $  1.84_{- 1.50}^{+ 2.18}$ & $  1.72_{- 0.34}^{+ 0.34}$ & $  0.59_{- 0.13}^{+ 0.14}$ & $ 24.92_{- 1.25}^{+ 1.19}$ & $ 24.57_{- 1.24}^{+ 1.17}$ & $    0.51_{- 0.12}^{+ 0.08}$\\
PSZ2 G239.93-39.97 &  &  & & $ 0.580 $ & $  71.6966 $ & $ -37.0625 $ & $  1.44 $ & $           34,           23 $  & 0679181001, 0693661201 & $         1022_{-          14}^{+          15}$ & $  5.73_{- 0.23}^{+ 0.25}$ & $  2.31_{- 1.90}^{+ 2.71}$ & $  2.49_{- 0.45}^{+ 0.45}$ & $  2.25_{- 0.13}^{+ 0.14}$ & $ 20.65_{- 0.55}^{+ 0.54}$ & $ 20.47_{- 0.54}^{+ 0.54}$ & $   -0.95_{- 0.10}^{+ 0.13}$\\
PSZ2 G254.64-45.20 & SPT-CLJ0417-4748 &  & & $ 0.581 $ & $  64.3464 $ & $ -47.8134 $ & $  1.34 $ & $           20,            9 $  & 0700182401 & $          974_{-          18}^{+          17}$ & $  4.97_{- 0.27}^{+ 0.27}$ & $  1.95_{- 1.57}^{+ 2.34}$ & $  2.54_{- 0.52}^{+ 0.52}$ & $  0.74_{- 0.11}^{+ 0.11}$ & $ 48.76_{- 0.88}^{+ 0.88}$ & $ 47.76_{- 0.86}^{+ 0.86}$ & $    1.97_{- 0.08}^{+ 0.09}$\\
PSZ2 G144.83+25.11 &  &  & MACSJ20647.7+7015& $ 0.584 $ & $ 101.9590 $ & $  70.2481 $ & $  5.40 $ & $           72,           46 $  & 0551850401,  0551851301 & $         1130_{-          10}^{+          10}$ & $  7.78_{- 0.20}^{+ 0.20}$ & $  3.51_{- 3.12}^{+ 3.89}$ & $  2.22_{- 0.25}^{+ 0.25}$ & $  0.33_{- 0.04}^{+ 0.04}$ & $ 30.83_{- 0.34}^{+ 0.32}$ & $ 30.66_{- 0.34}^{+ 0.32}$ & $    1.02_{- 0.03}^{+ 0.03}$\\
PSZ2 G045.32-38.46 &  &  & MACSJ2129.4-0741& $ 0.589 $ & $ 322.3591 $ & $  -7.6913 $ & $  4.32 $ & $            9,            3 $  & 0700182001 & $         1109_{-          34}^{+          32}$ & $  7.39_{- 0.65}^{+ 0.66}$ & $  3.08_{- 2.46}^{+ 3.70}$ & $  2.40_{- 0.53}^{+ 0.53}$ & $  2.46_{- 0.20}^{+ 0.22}$ & $ 30.05_{- 1.19}^{+ 1.11}$ & $ 29.85_{- 1.18}^{+ 1.10}$ & $   -0.34_{- 0.17}^{+ 0.15}$\\
PSZ2 G070.89+49.26 &  &  & WHL J155625.2+444042& $ 0.602 $ & $ 239.1098 $ & $  44.6772 $ & $  1.23 $ & $           47,           25 $  & 0693661301 & $          970_{-          14}^{+          13}$ & $  5.02_{- 0.21}^{+ 0.20}$ & $  2.28_{- 2.00}^{+ 2.56}$ & $  2.20_{- 0.29}^{+ 0.29}$ & $  0.25_{- 0.08}^{+ 0.08}$ & $ 14.68_{- 0.50}^{+ 0.47}$ & $ 14.62_{- 0.49}^{+ 0.47}$ & $   -0.52_{- 0.07}^{+ 0.07}$\\
PSZ2 G045.87+57.70 &  &  & WHL J151820.6+292740& $ 0.609 $ & $ 229.5866 $ & $  29.4603 $ & $  2.12 $ & $           23,           15 $  & 0693661101 & $         1016_{-          13}^{+          13}$ & $  5.82_{- 0.22}^{+ 0.22}$ & $  2.13_{- 1.84}^{+ 2.43}$ & $  2.73_{- 0.39}^{+ 0.39}$ & $  0.77_{- 0.06}^{+ 0.06}$ & $ 39.38_{- 0.61}^{+ 0.62}$ & $ 37.97_{- 0.59}^{+ 0.60}$ & $    1.36_{- 0.05}^{+ 0.05}$\\
PSZ2 G073.31+67.52 &  &  & WHL J215.168+39.91& $ 0.609 $ & $ 215.1709 $ & $  39.9187 $ & $  0.82 $ & $           26,           19 $  & 0693661001 & $         1035_{-          14}^{+          14}$ & $  6.15_{- 0.25}^{+ 0.26}$ & $  1.45_{- 1.15}^{+ 1.75}$ & $  4.25_{- 0.90}^{+ 0.89}$ & $  0.33_{- 0.10}^{+ 0.08}$ & $ 18.95_{- 0.53}^{+ 0.52}$ & $ 18.71_{- 0.52}^{+ 0.51}$ & $   -0.02_{- 0.07}^{+ 0.07}$\\
PSZ2 G099.86+58.45 &  &  & WHL J213.697+54.78& $ 0.615 $ & $ 213.6952 $ & $  54.7840 $ & $  1.50 $ & $           21,            9 $  & 0693660601, 0693662701, 0723780301 & $         1082_{-          22}^{+          21}$ & $  7.09_{- 0.42}^{+ 0.42}$ & $  2.45_{- 1.92}^{+ 2.99}$ & $  2.89_{- 0.65}^{+ 0.65}$ & $  1.43_{- 0.17}^{+ 0.16}$ & $ 17.15_{- 0.65}^{+ 0.59}$ & $ 17.02_{- 0.64}^{+ 0.58}$ & $   -0.78_{- 0.12}^{+ 0.10}$\\
PSZ2 G193.31-46.13 &  &  & & $ 0.634 $ & $  53.9644 $ & $  -6.9758 $ & $  4.15 $ & $           67,           48 $  & 0658200401, 0693661501 & $          940_{-          15}^{+          14}$ & $  4.76_{- 0.23}^{+ 0.22}$ & $  0.00_{- 0.00}^{+ 0.00}$ & $  0.00_{- 0.00}^{+ 0.00}$ & $ 11.95_{- 0.20}^{+ 0.21}$ & $  7.06_{- 0.20}^{+ 0.19}$ & $  6.63_{- 0.19}^{+ 0.18}$ & $  -10.04_{- 0.13}^{+ 0.14}$\\
PLCK G147.3-16.6 &  &  & & $ 0.645 $ & $  44.1056 $ & $  40.2885 $ & $  8.29 $ & $           59,           39 $  & 0679181301, 0693661601 & $         1033_{-          15}^{+          16}$ & $  6.39_{- 0.28}^{+ 0.30}$ & $  3.13_{- 2.65}^{+ 3.60}$ & $  2.04_{- 0.32}^{+ 0.32}$ & $  2.59_{- 0.12}^{+ 0.12}$ & $ 10.78_{- 0.30}^{+ 0.30}$ & $ 10.64_{- 0.29}^{+ 0.29}$ & $   -2.46_{- 0.08}^{+ 0.07}$\\
PLCK G260.7-26.3 & SPT-CLJ0616-5227 & ACT-CL J0616-5227 & & $ 0.680 $ & $  94.1429 $ & $ -52.4518 $ & $  4.25 $ & $           35,           20 $  & 0693662301 & $          910_{-          22}^{+          22}$ & $  4.56_{- 0.32}^{+ 0.34}$ & $  1.70_{- 1.46}^{+ 1.94}$ & $  2.68_{- 0.42}^{+ 0.43}$ & $  2.90_{- 0.19}^{+ 0.20}$ & $ 28.90_{- 0.91}^{+ 0.90}$ & $ 28.52_{- 0.90}^{+ 0.88}$ & $   -0.75_{- 0.14}^{+ 0.14}$\\
PSZ2 G219.89-34.39 &  &  & & $ 0.700 $ & $  73.6894 $ & $ -20.2851 $ & $  3.25 $ & $           51,           34 $  & 0679180501, 0693660301 & $         1030_{-          15}^{+          16}$ & $  6.77_{- 0.29}^{+ 0.33}$ & $  3.27_{- 2.69}^{+ 3.84}$ & $  2.07_{- 0.37}^{+ 0.38}$ & $  3.29_{- 0.13}^{+ 0.12}$ & $ 18.70_{- 0.51}^{+ 0.47}$ & $ 18.50_{- 0.51}^{+ 0.46}$ & $   -1.99_{- 0.10}^{+ 0.10}$\\
PSZ2 G208.61-74.39 &  &  & & $ 0.711 $ & $  30.0695 $ & $ -24.9132 $ & $  1.37 $ & $           47,           28 $  & 0693662901, 0723780601 & $          941_{-          14}^{+          14}$ & $  5.23_{- 0.23}^{+ 0.23}$ & $  1.83_{- 1.57}^{+ 2.10}$ & $  2.85_{- 0.43}^{+ 0.43}$ & $  1.93_{- 0.14}^{+ 0.14}$ & $ 15.97_{- 0.58}^{+ 0.57}$ & $ 15.64_{- 0.56}^{+ 0.56}$ & $   -1.31_{- 0.12}^{+ 0.12}$\\
PSZ2 G352.05-24.01 &  &  & & $ 0.798 $ & $ 290.2490 $ & $ -45.8500 $ & $  6.03 $ & $           52,           28 $  & 0679180201, 0693660401 & $          925_{-          17}^{+          18}$ & $  5.50_{- 0.31}^{+ 0.32}$ & $  2.23_{- 1.91}^{+ 2.55}$ & $  2.46_{- 0.38}^{+ 0.38}$ & $  0.31_{- 0.07}^{+ 0.07}$ & $ 28.21_{- 0.86}^{+ 0.85}$ & $ 28.00_{- 0.85}^{+ 0.84}$ & $    0.86_{- 0.08}^{+ 0.05}$\\
\hline
 & SPT-CLJ2146-4633 &  & & $ 0.933 $ & $ 326.6447 $ & $ -46.5475 $ & $  1.64 $ & $          153,          102 $  & 0744400501, 0744401301 & $          728_{-          11}^{+          10}$ & $  3.15_{- 0.14}^{+ 0.13}$ & $  0.34_{- 0.24}^{+ 0.43}$ & $  9.39_{- 2.73}^{+ 2.70}$ & $  2.49_{- 0.17}^{+ 0.17}$ & $  9.72_{- 1.09}^{+ 1.14}$ & $  9.44_{- 1.06}^{+ 1.11}$ & $ -1.65_{- 0.17}^{+ 0.12}$\\
PSZ2 G266.54-27.31 & SPT-CLJ0615-5746 &  & & $ 0.972 $ & $  93.9660 $ & $ -57.7796 $ & $  4.32 $ & $           12,            3 $  & 0658200101 & $          993_{-          14}^{+          14}$ & $  8.38_{- 0.36}^{+ 0.35}$ & $  3.18_{- 2.66}^{+ 3.69}$ & $  2.63_{- 0.44}^{+ 0.44}$ & $  0.68_{- 0.18}^{+ 0.18}$ & $ 34.06_{- 0.42}^{+ 0.38}$ & $ 33.43_{- 0.41}^{+ 0.37}$ & $  1.13_{- 0.16}^{+ 0.15}$\\
 & SPT-CLJ2341-5119 &  & & $ 1.003 $ & $ 355.3010 $ & $ -51.3286 $ & $  1.21 $ & $           91,           45 $  & 0744400401, 0763670201 & $          777_{-          11}^{+          11}$ & $  4.16_{- 0.17}^{+ 0.17}$ & $  1.76_{- 1.42}^{+ 2.10}$ & $  2.37_{- 0.47}^{+ 0.47}$ & $  1.96_{- 0.16}^{+ 0.16}$ & $ 27.19_{- 1.29}^{+ 1.17}$ & $ 26.82_{- 1.27}^{+ 1.15}$ & $ -0.20_{- 0.11}^{+ 0.19}$\\
 & SPT-CLJ0546-5345 & ACT-CL J0546-5345 & & $ 1.066 $ & $  86.6551 $ & $ -53.7596 $ & $  6.79 $ & $          127,          113 $  & 0744400201, 0744400301 & $          762_{-          10}^{+          10}$ & $  4.21_{- 0.16}^{+ 0.18}$ & $  2.81_{- 2.38}^{+ 3.24}$ & $  1.50_{- 0.23}^{+ 0.24}$ & $  1.16_{- 0.11}^{+ 0.11}$ & $ 20.46_{- 1.08}^{+ 1.57}$ & $ 19.90_{- 1.05}^{+ 1.53}$ & $  0.04_{- 0.12}^{+ 0.10}$\\
 & SPT-CLJ2106-5844 &  & & $ 1.132 $ & $ 316.5221 $ & $ -58.7421 $ & $  4.33 $ & $           26,           16 $  & 0763670301 & $          880_{-          19}^{+          18}$ & $  7.00_{- 0.43}^{+ 0.43}$ & $  0.67_{- 0.47}^{+ 0.88}$ & $ 10.42_{- 3.26}^{+ 3.24}$ & $  1.55_{- 0.21}^{+ 0.21}$ & $ 15.56_{- 0.70}^{+ 0.66}$ & $ 15.31_{- 0.69}^{+ 0.65}$ & $ -0.58_{- 0.17}^{+ 0.14}$\\

\hline
\end{tabular}
}
\end{center}
\end{sidewaystable*}
\end{document}